\documentclass[onecolumn, 11pt, amsmath, amssymb, aps, prd]{revtex4-2}

\usepackage[utf8]{inputenc}
\usepackage[english]{babel}
\usepackage{hyperref}
\usepackage{enumitem}
\usepackage{relsize}
\usepackage{xcolor}
\usepackage{amsmath}
\usepackage[left=2.75cm,right=2.75cm,top=1.75cm,bottom=1.75cm,includeheadfoot]{geometry}

\newcommand{\ham}{\mathcal{H}}
\newcommand{\diff}{\mathcal{D}}
\newcommand{\erad}{E^r}
\newcommand{\ephi}{E^\varphi}
\newcommand{\krad}{K_r}
\newcommand{\kang}{K_\varphi}
\newcommand{\pphi}{P_\phi}

\newcommand{\hol}{h_2}
\newcommand{\holbar}{h_1}

\newcommand{\lapse}{N}
\newcommand{\shift}{N^r}

\begin{document}

\title{{{Holonomy and inverse-triad corrections in spherical models coupled to matter}}}

\author{{{Asier Alonso-Bardaji\,}}}
 \email{asier.alonso@ehu.eus}

\author{{{David Brizuela\,}}}
 \email{david.brizuela@ehu.eus}
 
\affiliation{{\small{Fisika Saila, Universidad del Pa\'is Vasco/Euskal Herriko Unibertsitatea (UPV/EHU), Barrio Sarriena s/n, 48940 Leioa, Spain}}}

\begin{abstract}
Loop quantum gravity introduces two characteristic modifications in the classical constraints of general relativity: the holonomy and inverse-triad corrections. In this paper, a systematic construction of anomaly-free effective constraints encoding such corrections is developed for spherically symmetric spacetimes. The starting point of the analysis is a generic Hamiltonian constraint where free functions of the triad and curvature components as well as non-minimal couplings between geometric and matter degrees of freedom are considered. Then, the requirement of anomaly freedom is imposed in order to obtain a modified Hamiltonian that forms a first-class algebra. In this way, we construct a family of consistent deformations of spherical general relativity, which generalizes previous results in the literature. The discussed derivation is implemented for vacuum as well as for two matter models: dust and scalar field. Nonetheless, only the deformed vacuum model admits free functions of the connection components. Therefore, under the present assumptions, we conclude that holonomy corrections are not allowed in the presence of these matter fields.
\end{abstract}

\maketitle

\section{Introduction}\label{sec:intro}

One of the main aspects a quantum theory of gravity must face
is that regarding the singularities of general relativity. 
The discrete nature of spacetime predicted by loop quantum 
gravity may provide an answer to this 
problem \cite{thiemannbook}. In fact, symmetry-reduced homogeneous models have 
been widely studied, particularly those related to cosmology, yielding classical singularity 
resolution via a cosmological bounce \cite{lqcreport,lqcbrief,lqgresumen}.

However, the extreme energy scales at which quantum-gravity phenomena become relevant
make experimental evidence still unreachable and predictions of any quantum-gravity
theory seem hard to test. Nevertheless, a general expectation is that these effects
leave observable trails at lower energies, which could be studied via effective theories.
In fact, in the context of loop quantum cosmology, effective schemes provide
an accurate description of the evolution when compared to the full quantum dynamics \cite{lqcreport,lqcbrief}.
In addition, these effective theories have been widely used to study the interior of spherical black holes,
which are described by homogeneous but anisotropic Kantowski-Sachs spaces. The literature presents
a wide variety of predictions \cite{AshtekarBojowald,gambini&pullin,corichibh,Barrau:2018rts,pertashtekar,Ashtekarkruskal,tecotl,bmm,Bouhmadi2020,olmedobh}; for instance, the
possibility of a quantum transition from a black hole into a white hole,
resembling the cosmological bounce.

The usual approach to obtain effective equations
is to modify the Hamiltonian in such a way that expected quantum effects from loop quantum gravity are included. Among these modifications, we find two main branches: classical divergences when the volume of a region tends to zero are regularized and encoded in the so-called inverse-triad modifications, whereas holonomy corrections are directly related to the spacetime discreteness predicted by the theory.
In homogeneous models, as the diffeomorphism constraint vanishes
off-shell, one can include by hand a wide variety of modifications.
However, the construction of a consistent effective theory becomes rather challenging when non-homogeneous scenarios are considered. In these cases, the diffeomorphism constraint is no longer vanishing and one finds that the quantized notion of spacetime collides
with the continuous diffeomorphism symmetry of general relativity.

Several different attempts to study inhomogeneous models can be found in
the literature. For instance, Gowdy models have been analyzed by means of the procedure named
hybrid quantization \cite{MartinBenito:2008ej}. This approach splits the problem in two: a Bianchi I model describing the homogeneous background spacetime, quantized following the loop quantum cosmology program, and a Fock quantization of the remaining inhomogeneities.
This scheme has also been applied to cosmological perturbations
\cite{FernandezMendez:2012vi, Fernandez-Mendez:2013jqa, 1407.0998, PhysRevD.93.104025}. In the same context, one must also mention the dressed-metric approach \cite{PhysRevD.87.043507,Agullo_2013, Agullo_2017}. Both methods make similar assumptions and provide a
completely quantized theory to analyze cosmological perturbations.
However, it is not clear whether these frameworks respect the general covariance
of the theory \cite{PhysRevD.102.023532}.
In fact, this is the main guideline of the consistent constraint deformation approach.
This alternative formalism constructs effective theories demanding that the modified
constraints form an anomaly-free algebra. References 
\cite{anomalpert,pert1,pert2,pertnew,pertnew2,pertholonomy}
implement this approach for cosmological perturbations.

Spherically symmetric models are of particular relevance, as they would be a first step to 
describe black holes and gravitational collapse. 
In this context, it was found that the Abelianization of the constraints could lead to a consistent quantization \cite{PhysRevLett.110.211301},
even under the presence of a scalar matter field \cite{Gambini_2014}.
Nonetheless, this approach suffers equally from the lack of covariance,
as analyzed in \cite{bojo2}.
On the other hand, the consistent constraint deformation approach has already been applied
to several spherical models, both in vacuum and coupled to different
matter fields \cite{ltb,tesisreyes,tibrewala,bojo2,algebroids,bojo1,nodiff},
leading to promising results.
In particular, the effective line element corresponding
to the holonomy-corrected constraints might present a
signature change when approaching the classical singularity.
From this standpoint, some interior region of the black hole would be Euclidean,
and the usual notions of causality and time evolution would not apply.
In particular, in \cite{bojobh}, it is argued that this prediction rules
out the possibility of bouncing black holes.
Unfortunately, most studies based on this methodology acknowledge that holonomy modifications
can only be implemented in vacuum.
The requirement of anomaly freedom in the presence of matter with local degrees of
freedom restricts even more feasible modifications, leaving no room for holonomy corrections
\cite{bojo1}.

Therefore, the goal of this paper is to perform a systematic analysis of possible
modifications to classical constraints of general relativity in spherical
symmetry. We will consider the vacuum case as well as gravity coupled to
two different matter models: dust and scalar field. A key difference between
the present and previous studies will be the possibility for non-minimal couplings
between matter and geometric degrees of freedom as well as for
polymerization of the matter variables. Nonetheless, in agreement with
previous works, we will find that only the vacuum model admits holonomy
corrections.
  
  The rest of the paper is organized as follows. Section~\ref{sec:algebra} includes a brief summary of the constraint structure of classical general relativity~(\ref{sec:class}) and introduces the main features of inverse-triad and holonomy corrections~(\ref{sec:loops}). In  section~\ref{sec:polgrav}, we compute the Poisson brackets between modified constraints, assuming a completely general kinetic part in the Hamiltonian for the vacuum model. In sections \ref{sec:scalarfield} and \ref{sec:dust}, respectively, we will extend our analysis to include a scalar field and dust with possible non-minimal couplings between matter and geometry. Finally, section~\ref{sec:conclusions} summarizes the main results and presents the conclusions of our study.

\section{Constraint algebra}\label{sec:algebra}

\subsection{Classical theory}\label{sec:class}

    The canonical formalism of general relativity shows that this theory is a completely constrained dynamical system. The total Hamiltonian is a linear combination of the so-called Hamiltonian constraint and diffeomorphism constraint, with the coefficients given by the lapse function ($\lapse$) and the shift vector, respectively. Since we will be assuming spherical symmetry, the angular components of the diffeomorphism constraint will vanish off-shell and the only relevant part of the shift vector will be its radial component ($\shift$). 
      
  Following the usual approach in loop quantum gravity, the basic variables will be the $U(1)$-invariant components of the spherically symmetric triad, $\erad$ and $\ephi$. Their canonically conjugate momenta will be given by the elements of the Ashtekar $su(2)$ connection, which can be written in terms of the extrinsic curvature components $\krad$ and $\kang$. Setting the Newton constant to one, $G=1$, the symplectic structure of the phase space is given by the Poisson brackets,
\vspace{-0.05cm} 
  \begin{equation}\label{eq:vars}
    \lbrace K_r(x),E^r(y)\rbrace = \lbrace\kang(x),\ephi(y)\rbrace = \delta(x-y)\:.
  \end{equation}
  In terms of these variables, the vacuum Hamiltonian and diffeomorphism
  constraints of classical general relativity read respectively as follows,
  \begin{align}
    \label{eq:classhamg}
   \ham_{g} &:= \sqrt{\erad}\Big({{\Gamma_{\varphi}}}\!'-2\krad\kang\Big)\!-\!\frac{\ephi}{2\sqrt{\erad}}\Big(1+{\kang}\!^2-{\Gamma_{\varphi}\!}^2\Big),\\\label{eq:classdiffg}
    \diff_{g} &:= (\kang)'\ephi-(\erad)'\krad\:,
  \end{align}
  where $\Gamma_{\!\varphi}:=(\erad)'/(2\ephi)$ is the angular component of the spin connection.
  
  In the canonical formalism, the covariance of general relativity is not explicit, but it can be
  read from the \emph{hypersurface deformation algebra},
  \begin{subequations}\label{eq:hda}
  \begin{align}
     \{ D[\shift_1] , D[\shift_2] \} &= D[\shift_1{\shift_2}'-{\shift_1}'\shift_2]\:,\label{eq:classdd}\\
     \{ D[\shift] , H[\lapse] \} &= H[\shift\lapse']\:,\label{eq:classdh}\\
     \{ H[\lapse_1] , H[\lapse_2] \} &= D[q^{rr}(\lapse_1{\lapse_2}\!'-{\lapse_1}\!'\lapse_2)]\label{eq:classhh}\:,
  \end{align} 
  \end{subequations}
where the smeared form of the constraints has been introduced as {\small $D[\shift]:=\int dr \shift\diff$} and {\small $H[\lapse]:=\int dr\lapse\ham$}, and the prime denotes the derivative with respect to the radial variable $r$. Note that, in the last bracket, the component $q^{rr}$ of the inverse three-metric appears, which can be obtained from the spherically symmetric spatial line element,
  \begin{equation}\label{eq:3metric}
    {d\sigma}^2 = \frac{(\ephi)^2}{\erad}{dr}^2 +\erad {d\Omega}^2\:.
  \end{equation}

\subsection{Corrections motivated by loop quantum gravity}\label{sec:loops}

   In this section, the two different types of corrections motivated by the theory of loop quantum gravity are reviewed: the so-called inverse-triad and holonomy corrections. 
 
\subsubsection{Inverse-triad corrections}\label{sec:triads}
  
    In loop quantum gravity, the operators built from triad components contain zero in their discrete spectra. Therefore, they cannot be inverted. Nonetheless, a regularization procedure is followed to obtain well-defined operators \cite{thiemann,bojovol}. The expectation values of these regularized operators mimic the classical behavior at large scales but present deviations from classical divergences at small volumes. In the context of effective theories, inverse-triad corrections are introduced to account for the regularization carried out in the full theory.
      
  In principle, only terms with explicit triad terms in the denominator should be held to corrections. However,
  in the symmetry reduction process that leads to expression \eqref{eq:classhamg}, some triad factors have been simplified coming from the volume element.
  At this level, we will not discuss the implications of correcting just the volume element or individual terms in the Hamiltonian. We will introduce two independent multiplicative functions, $\alpha_1$ and $\alpha_2$,
  \begin{align}\label{eq:triadham}
    {\ham}_{g}^{(t)} :=
    \alpha_1(\erad)\sqrt{\erad}\Big({\Gamma_{\!\varphi}}'-2\krad\kang\Big)-\alpha_2(\erad)\frac{\ephi}{2\sqrt{\erad}}\Big(1+{\kang}\!^2-{\Gamma_{\!\varphi}\!}^2\Big)\:,
  \end{align}
  so that when both tend to one, we recover the classical Hamiltonian constraint \eqref{eq:classhamg}. If the origin of the corrections was the volume element, then both correction functions would be equal $\alpha_1=\alpha_2$. 
  
  One can check that this deformed Hamiltonian and the classical diffeomorphism constraint \eqref{eq:classdiffg} do close algebra. More precisely, the algebraic relations \eqref{eq:classdd} and \eqref{eq:classdh} remain unmodified, but the third bracket \eqref{eq:classhh} acquires a deformation given by the function ${\alpha_1}$:
  \begin{equation}\label{eq:triadhh}
    \{ {H}_{g}^{(t)}[\lapse_1] , {H}_{g}^{(t)}[\lapse_2] \} = D_g\big[{\alpha_1^2}\,q^{rr}(\lapse_1{\lapse_2}\!'-{\lapse_1}\!'\lapse_2)\big]\:,
  \end{equation}
  where ${H}_{g}^{(t)}[\lapse]$ is the smeared form of ${\cal H}_{g}^{(t)}$. Note that if only terms with explicit powers of inverse triads were modified in \eqref{eq:classhamg}, $\alpha_1$ would be one and the whole algebraic structure would remain undeformed.

\subsubsection{Holonomy modifications}\label{sec:holonomies}
  
    The operator associated to the connection is not well
  defined in loop quantum gravity. Alternatively, one considers holonomies, the exponential form of parallel-transported connections, which do have a definite operator counterpart. In order to
  construct effective theories, the general idea is that  connection components should be replaced with periodic functions. This procedure is sometimes referred to as polymerization.
  
  In the present approach, holonomy corrections will involve the replacement of the extrinsic curvature variables by generic functions. The requisite of a closed algebra will impose conditions over these modifications.
  
  For instance, following the standard procedure found in the literature \cite{tesisreyes,tibrewala,bojo2}, let us replace the terms $\kang$ and ${\kang}\!^2$ in expression \eqref{eq:classhamg} by $f_1(\kang)$ and  $f_2(\kang)$, respectively, and define the holonomy-modified Hamiltonian constraint,
  \begin{equation}\label{eq:holonomyham}
      \ham_{g}^{(h)} := \sqrt{\erad}\Big({\Gamma_{\!\varphi}}'-2\krad f_1\Big)-\frac{\ephi}{2\sqrt{\erad}}\Big(1+f_2-{\Gamma_{\!\varphi}\!}^2\Big)\:,
  \end{equation}
along with its smeared version ${H}_{g}^{(h)}[\lapse]:=\int dr \lapse \ham_{g}^{(h)}$.
 
 Considering this deformed Hamiltonian in combination with the classical diffeomorphism constraint \eqref{eq:classdiffg}, one can check that the first two brackets \eqref{eq:classdd} and \eqref{eq:classdh} of the hypersurface deformation algebra remain unmodified, while the third one generates an anomalous term:
  \begin{align}\label{eq:holonomyhh}
    \{ {H}_{g}^{(h)}[\lapse_1] , {H}_{g}^{(h)}[\lapse_2] \} &= D_g\!\left[\frac{\partial f_1}{\partial\kang} q^{rr}(\lapse_1{\lapse_2}\!'-{\lapse_1}\!' \lapse_2)\right]\nonumber\\&+ \frac{(\erad)'}{4\ephi}(\lapse_1{\lapse_2}\!'-{\lapse_1}\!'\lapse_2)\left(2f_1-\frac{\partial f_2}{\partial\kang}\right).
  \end{align}
  In order to cancel the anomaly, one needs to enforce relation $2f_1=\partial f_2/\partial\kang$. Thus, both functions are not independent and, following the motivation mentioned above, they are usually chosen to have a sinusoidal form. For example, $f_1=\sin(\mu \kang)/\mu$, where the parameter $\mu$ is assumed to be related to the minimum area predicted by the theory. In the limit 
  $\mu\rightarrow 0$, the Hamiltonian \eqref{eq:holonomyham} recovers its classical form \eqref{eq:classhamg}.
   
  Finally, let us emphasize that the two types of corrections introduced in this section are compatible. That is, one can consider the deformed Hamiltonian constraint,
  \begin{align}\label{hamiltonianht}
      \ham_{g}^{(ht)} := \alpha_1(\erad)\sqrt{\erad}\left({\Gamma_{\!\varphi}}'-2\krad \frac{\partial f(\kang)}{\partial\kang}\right)-\alpha_2(\erad)\frac{\ephi}{2\sqrt{\erad}}\Big(1+2f(\kang)-{\Gamma_{\varphi}\!}^2\Big)\:,
  \end{align}
  with generic functions $\alpha_1$, $\alpha_2$ and $f$. This Hamiltonian closes algebra along with the classical diffeomorphism constraint \eqref{eq:classdiffg}. The brackets \eqref{eq:classdd} and \eqref{eq:classdh} remain unchanged, whereas the one between modified Hamiltonian constraints presents a deformation equal to the product of corrections in the right-hand sides of \eqref{eq:triadhh} and \eqref{eq:holonomyhh}:
  \begin{align}\label{eq:hthh}
     \big\{ {H}_{g}^{(ht)}[\lapse_1] , {H}_{g}^{(ht)}[\lapse_2] \big\} = D_g\bigg[\alpha_1^2\frac{\partial^2f}{\partial\kang^2}\,q^{rr}
     \big(\lapse_1{\lapse_2}\!'-{\lapse_1}\!' \lapse_2\big)\bigg]\:,
  \end{align}
  with the smeared form $H_g^{(ht)}:=\int dr \lapse \ham_g^{(ht)}$.
    
\section{Vacuum}\label{sec:polgrav}
  
    Before studying the possibility of constructing effective theories with inverse-triad and holonomy corrections in the presence of matter fields, let us first try to generalize the known results for vacuum summarized in the previous section. For such a purpose, we will consider the following deformed Hamiltonian constraint,
  \begin{align}\label{eq:polhamg}
      \widetilde{\ham}_g \!=\!
    f_g(\erad,\ephi,\krad,\kang)+\eta_1(\erad,\ephi)\sqrt{\erad}{\Gamma_{\!\varphi}}'-\eta_2(\erad,\ephi)\frac{\ephi}{2\sqrt{\erad}}+\eta_3(\erad,\ephi)\frac{\ephi{\Gamma_{\!\varphi}\!}^2}{2\sqrt{\erad}}.
  \end{align}
  Note that the usual kinetic part (all terms with an explicit factor of $\krad$
  or $\kang$) of the classical Hamiltonian constraint \eqref{eq:classhamg} has been
  replaced with a generic function $f_g$, which is assumed to depend on all canonical variables.
  This function will encode the possible holonomy corrections of the model
  as well as inverse-triad modifications. In this way, the holonomy corrections might also depend on triad components, as
  in the improved-dynamics scheme of loop quantum cosmology \cite{0607039}.
  Additionally, we have included three free functions $\eta_1$, $\eta_2$, and $\eta_3$, with dependence on both triad components $\erad$ and $\ephi$,  in front of each remaining term to account for additional inverse-triad corrections.
   In fact, we could expand the derivative of the spin-connection ${\Gamma_{\!\varphi}}'$ in two terms depending on the derivatives of the triad components and include a correction multiplying each of them. Instead of just one function ($\eta_1$), this would  lead to two independent corrections but with the same final results, since anomaly
  freedom would demand these two functions to be equal. Therefore,
  without loss of generality, we will just consider the correction
  function $\eta_1$.

  Concerning the diffeomorphism constraint, in this section we will keep it unmodified
  and its classical form, given in \eqref{eq:classdiffg}, will be considered.
  The full theory of loop quantum gravity does not modify the diffeomorphism constraint
  and thus one does not expect it to be modified
  at an effective level.
  However, in App.~\ref{app:moddiff},
  we will consider generic deformations of the diffeomorphism constraint,
  along with the form \eqref{eq:polhamg} for the Hamiltonian. Following the same methodology as in this section, we will explicitly show that (up to canonical transformations) only a global rescaling of the
  diffeomorphism constraint is allowed, leading to the same modified Hamiltonian that will
  be obtained in this section (expression \eqref{eq:polhamgrav} below). Results in App.~\ref{app:moddiff} generalize previous studies that point out that the requirement of anomaly freedom only admits trivial modifications of the diffeomorphism constraint \cite{nodiff}.
  
  Regarding the constraint algebra, it is clear that,
  since here we did not changed the diffeomorphism constraint,
  relation \eqref{eq:classdd} remains unmodified. On the other hand, the bracket between the classical diffeomorphism \eqref{eq:classdiffg} and the modified Hamiltonian constraints \eqref{eq:polhamg} gives rise, up to irrelevant global factors, to the following anomalous term,
    \begin{align}\label{eq:anomalgdh}
      \mathcal{A}_g^{DH} &:=8\sqrt{\erad}\Bigg[f_g-\krad\frac{\partial f_g}{\partial\krad}-\eta_1\ephi\frac{\partial( f_g/ \eta_1)}{\partial\ephi}\Bigg]
      +{\eta_1}\big[(\erad)'\big]^2\frac{\partial (\eta_3/\eta_1)}{\partial\ephi}
    \nonumber\\&- {4}{\eta_1}(\ephi)^2\frac{\partial (\eta_2/\eta_1)}{\partial\ephi}\:,
  \end{align}
  whereas the bracket of the modified Hamiltonian with itself leads
  to the anomaly:
  \begin{align}\label{eq:anomalghh}  
    \mathcal{A}_g^{HH} &:=(\erad)'\Bigg[\left(\eta_1-\eta_3\right)\frac{\ephi}{2}\frac{\partial f_g}{\partial\krad}-{\eta_1}\!^2\erad\ephi\frac{\partial^2 (f_g/\eta_1)}{\partial\krad\partial\erad}+\eta_1\erad\left(\frac{\partial f_g}{\partial\kang}-\krad\frac{\partial^2 f_g}{\partial\krad\partial\kang}\right)\!\! \Bigg]\nonumber\\
    &+ 
    (\ephi)'\erad\ephi\eta_1^2\frac{\partial^2(f_g/\eta_1)}{\partial\ephi\partial\krad}-(\krad)'\erad\ephi \eta_1\frac{\partial^2f_g}{\partial\krad^2}   \:. 
  \end{align}
   Therefore, the condition that the constraints $\widetilde{\cal H}_g$ and ${\cal D}_g$ form a closed algebra is translated to the vanishing of all the above terms:
\begin{align}
  &\label{eq:eqanomalgdh}{\cal A}_g^{DH}=0\:,\\
  &
  \label{eq:eqanomalghh}{\cal A}_g^{HH}=0\:.
\end{align}
  Up to this point, there is some ambiguity in the generic functions defined in the modified constraint. In particular, the third term of \eqref{eq:polhamg}, the one that goes with the $\eta_2$ function, could have been absorbed in the definition of $f_g$ as it does not depend on the radial derivatives of the triad components. This means that, with no loss of generality, $\eta_2$ can be chosen as desired. Therefore, in order to cancel the last term of the anomaly \eqref{eq:anomalgdh}, we will demand,
\begin{equation}\label{eta2eta1}
\eta_2(\erad,\ephi)=\xi_2(\erad)\,\eta_1(\erad,\ephi)/\xi_1(\erad)\:,
\end{equation}
  $\xi_1$ and $\xi_2$ being free functions of the radial component of the triad.  Notice that the generic functions, and $\eta_1$ in particular, are assumed to be non-vanishing.
  
  None of the free functions that are contained in the anomaly equations depend on radial derivatives.
   Hence, each coefficient of a primed variable must vanish independently and the following five equations, alongside \eqref{eta2eta1}, are equivalent to the system \eqref{eq:eqanomalgdh}-\eqref{eq:eqanomalghh}:
  \begin{align}\label{eq1}
    0&=\frac{\partial (\eta_3/\eta_1)}{\partial\ephi}\:,\\\label{eq1bis}
    0&=f_g-\krad\frac{\partial f_g}{\partial\krad}-\eta_1\ephi\frac{\partial( f_g/ \eta_1)}{\partial\ephi}\:,\\
    \label{eq2}
    0&= \frac{\partial^2(f_g/\eta_1)}{\partial\ephi\partial\krad},\\\label{eq3}
    0&= (\eta_1-\eta_3)\frac{\ephi}{2}\frac{\partial f_g}{\partial\krad}-{\eta_1}\!^2\erad\ephi\frac{\partial^2 (f_g/\eta_1)}{\partial\krad\partial\erad}+\eta_1\erad\left(\frac{\partial f_g}{\partial\kang}-\krad\frac{\partial^2 f_g}{\partial\krad\partial\kang}\right)\!,\\\label{eq4}
    0&=\frac{\partial^2f_g}{\partial\krad^2}\:.
  \end{align}
  The general solution of equation \eqref{eq1} provides the following relation between the functions $\eta_1$ and $\eta_3$,
  \begin{equation}\label{eta3eta1}
     \eta_3(\erad,\ephi)=\xi_3(\erad)\,\eta_1(\erad,\ephi)/\xi_1(\erad),
  \end{equation}
  with $\xi_3$ a free function of the radial component of the triad. In addition,
  direct integration of equations \eqref{eq2} and \eqref{eq4} leads to the following
  form for the kinetic term of the Hamiltonian:
  \begin{align}\label{eq:fgform}
    f_g = \frac{\eta_1(\erad,\ephi)}{\xi_1(\erad)}\Big[\krad f_{1}(\erad,\kang)+f_{2}(\erad,\ephi,\kang)\Big],
  \end{align}
  where we have introduced two integration functions $f_1$ and $f_2$ along
  with their corresponding dependencies. In particular, note that
  the dependence on the radial component of the connection, $\krad$
  is already explicit and, therefore, holonomy corrections depending on $\krad$
  will not be allowed: the Hamiltonian must be linear in $\krad$.
  
 Replacing now \eqref{eq:fgform} in \eqref{eq1bis}, one obtains the relation,
  \begin{equation}\label{f0f1relation}
    f_2-\ephi \frac{\partial f_2}{\partial\ephi}=0\:,
  \end{equation}
  which implies that $f_2$ must be linear in $\ephi$.
  In summary, the expression
    \begin{align}\label{eq:fgfin}
      \!\!\!\!\!f_g\!=\!- \frac{\eta_1}{\xi_1}\Bigg[2\sqrt{\erad}\krad h_{1}(\erad\!,\kang) + \frac{\ephi}{2\sqrt{\erad}} h_{2}(\erad\!,\kang)\Bigg],
  \end{align}
  provides the most general function $f_g$, that obeys all conditions above.   For convenience, we have included some explicit numerical factors and $\sqrt{\erad}$
  terms, so that
  this function resembles the classical form of the kinetic part
  of the constraint when functions $h_1$ and $h_2$ take the values
  $\kang$ and ${\kang}\!^2$, respectively.

  In order to obtain an anomaly-free algebra, the only condition left
  is given by equation \eqref{eq3}. Inserting the forms \eqref{eta3eta1} and \eqref{eq:fgfin} in that relation, the following differential equation is obtained:
        \begin{equation}\label{eq:anomalg}
      \frac{1}{2}\frac{\partial \hol}{\partial\kang} =
     \left(\xi_3 -2\erad \frac{\partial\xi_1}{\partial\erad}\right)\!\frac{\holbar}{\xi_1} +2\erad \frac{\partial\holbar}{\partial\erad}.
    \end{equation}
     Collecting all the conditions we found for the correction functions, the deformed Hamiltonian constraint reads,
    \begin{align}\label{eq:polhamgrav}
    \widetilde{\ham}_g&= \eta(\erad,\ephi)\bigg[\sqrt{\erad}\Big(\xi_1(\erad){\Gamma_{\!\varphi}}'  -2\krad \holbar(\erad,\kang)\Big)\nonumber\\&
    -\frac{\ephi}{2\sqrt{\erad}}\Big(\xi_2(\erad) +\hol(\erad,\kang)-\xi_3(\erad){\Gamma_{\!\varphi}\!}^2\Big) \bigg],
  \end{align}
where we have defined the free global factor $\eta:=\eta_1/\xi_1$.
 Recall that all the modifications are not independent since the additional relation \eqref{eq:anomalg} must be obeyed.
 The classical Hamiltonian constraint is recovered when $\eta=\xi_1=\xi_2=\xi_3=1$, $h_1=\kang$ and $h_2=\kang^2$, which automatically satisfies the mentioned requisite.

  Regarding the physical interpretation of the modifications,
   functions $\xi_1$, $\xi_2$ and $\xi_3$ can be interpreted as inverse-triad corrections, whereas $h_1$ and $h_2$ would represent holonomy modifications. Since these last two have an arbitrary dependence on the radial component of the triad, they could also encode certain inverse-triad corrections.
  The factor $\eta$ is a trivial modification in the sense that one can always rescale
  the constraint by any global multiplicative function and the new rescaled constraint will provide an anomaly-free algebra. In addition, this global factor would not have any
  effect in local dynamics and might only alter global properties of the spacetime. Furthermore, since the Hamiltonian
  must be a scalar density, the global factor $\eta$ should be of weight zero. This would rule out
  the dependence of $\eta$ on 
  the angular component of the triad $\ephi$, for it being a scalar density.
  Nonetheless, in order to keep the discussion as general as possible, this dependence will be kept so that by incorporating additional densities a scalar object may be formed.

 The modified Hamiltonian constraint \eqref{eq:polhamgrav} is the main result of this
 section and provides a generalization of the already present results in the literature. In particular, expression \eqref{hamiltonianht} may be recovered
    just 
    taking $\eta=1$, $\xi_1=\alpha_1$, $\xi_2=\xi_3=\alpha_2$, $h_2=2\alpha_2 f$ and
    $h_1=\alpha_1\partial f/\partial\kang$. Note that, for these choices, relation
    \eqref{eq:anomalg} is automatically satisfied.
 
 The anomaly-free first-class algebra formed by this modified Hamiltonian and the classical diffeomorphism constraint is:  
  \begin{align}\label{eq:poldhgrav}
     \!\!\!\!\{ D_g[\shift] , \widetilde{H}_g[\lapse] \} \!&=\!\widetilde{H}_g\bigg[\!\shift\!\lapse'\!-\!(\shift)'\lapse\frac{\ephi}{\eta}\!\frac{\partial{\eta}}{\partial\ephi}\bigg],\\\label{eq:polhhgrav}
  \!\!\{ \widetilde{H}_g[\lapse_1] , \widetilde{H}_g[\lapse_2] \}\! &=\! D_g\bigg[\eta^2\xi_1\!\frac{\partial\holbar}{\partial\kang} q^{rr}(\lapse_1{\lapse_2}\!'-{\lapse_1}\!'\lapse_2)\!\bigg],
  \end{align}
    where $\widetilde{H}_g[\lapse]$ is the smeared version of $\widetilde{\ham}_g$. 
  The first bracket picks up corrections from the 
  global factor $\eta$. Nonetheless, the deformation
  of the bracket \eqref{eq:polhhgrav}
  might contain relevant information about quantum-gravity effects as one approaches the deep quantum regime, such as signature change \cite{bojobh,bojo1}.
  Only for the case $\eta^2\xi_1 h_1= \kang +g(\erad)$, 
  which would imply that $\eta$ is independent of $\ephi$, 
  one would recover the exact algebra of general relativity.
  
 The procedure we followed imposes the quite restrictive form \eqref{eq:polhamgrav} even if we started from a completely generic kinetic term in the Hamiltonian \eqref{eq:polhamg}, yielding some interesting features. First, we see that the only correction that depends on the angular component of the triad $\ephi$ is contained in the global factor $\eta$. Second, there are no free dependencies on the radial component of the connection $\krad$, which appear in the same way
  as in the classical Hamiltonian. Therefore, no holonomy modifications dependent on $\krad$ are allowed.

  The only possible holonomy corrections are encoded in $h_1$ and $h_2$, which are allowed to be scale-dependent through the radial component of the triad $\erad$. This is an interesting aspect of the model as it allows for some kind of improved-dynamics scheme, similar to
  the one considered in the context of loop quantum cosmology \cite{0607039}.
  In fact, there are already some examples in the literature with this kind of scale-dependent
  holonomy corrections.
  For instance, the anomaly equation \eqref{eq:anomalg} can be directly compared with results presented in \cite{tibrewala,bojo2} (which do not include additional inverse-triad
  corrections $\xi_1$, $\xi_2$, $\xi_3$). 
 Finally, in App.~\ref{app:leading} we analyze
  possible solutions for $h_1$ and $h_2$ that satisfy equation \eqref{eq:anomalg},  assuming
  a linear expansion of the functions and focusing on
  leading order terms.

  \subsubsection*{Singular solutions}
  
    For the sake of completeness, we would like to point out that, in order to write and solve the system of equations \eqref{eq:eqanomalgdh}--\eqref{eq:eqanomalghh}, all correction functions have been assumed to be non-vanishing. However, it is possible to
 find two families of \emph{singular} solutions (in the sense that they
  are not contained in the previous general case) that involve the vanishing of $\eta_1$:
 \begin{changemargin}{0.5cm}{0cm}
\begin{enumerate}[label=(\roman*)]
\item\label{c1} $\eta_1=0$ and $f_g=f_g(\erad,\ephi,\kang)$.
\item\label{c2} $\eta_1=0$ and $\eta_3=0$.
\end{enumerate}
\end{changemargin}

  Each of these modifications in the Hamiltonian provides a first-class algebra, without any further requirement
  on the remaining functions. Note that one can not directly substitute these values in the expressions above
  as $\eta_1$ appears dividing in the anomalous term \eqref{eq:anomalgdh}. In order to check that these are indeed consistent deformations, one needs to compute again the Poisson bracket \eqref{eq:poldhgrav}.
  
  Once $\eta_1=0$ is chosen, the only derivative term in the deformed Hamiltonian is $(\erad)'$. Next, one can consider either \ref{c1}, to eliminate its canonical conjugate variable $\krad$, or \ref{c2}, to remove
  all derivative terms in the Hamiltonian. The latter case corresponds to the homogeneous limit of the model which, following the Belinski–Khalatnikov–Lifshitz (BKL)
  conjecture \cite{bkl}, is expected to be relevant near the classical singularity.
  
  In contrast to the general case analyzed above, these singular solutions present a less constrained dependency on the connection variables. On the one hand, in \ref{c1}, although the dependence on $\krad$ is fixed, we find a complete freedom regarding $\kang$. On the other hand, case \ref{c2} brings complete freedom when choosing the dependence of
  the Hamiltonian constraint on both connection components, $\krad$ and $\kang$, allowing for any kind of holonomy correction.

\section{Matter fields}\label{sec:polmatter}
    
      The complexity of studying holonomy modifications
    in the presence of matter is widely recognized. In fact, there
    is a quite general argument that shows the impossibility of  consistently including holonomy corrections in the effective Hamiltonian 
    \cite{bojo2}.
    Let us consider a minimally-coupled matter model with a possibly deformed
    Hamiltonian constraint {\small{$\widetilde H_m[\lapse]$}}. The full smeared Hamiltonian is
    {\small{$\widetilde H[\lapse]=\widetilde H_g[\lapse]+\widetilde H_m[\lapse]$}}, with the vacuum
    Hamiltonian {\small{$\widetilde H_g[\lapse]$}} given by the smeared form of \eqref{eq:polhamgrav}. In order to define a first-class algebra, the bracket
   \begin{align}\label{fullhambracket}
      \big\{\widetilde H_g[\lapse_1]+\widetilde H_m[\lapse_1],\widetilde H_g[\lapse_2]+\widetilde H_m[\lapse_2]\big\},
    \end{align}
    should be expressed as a linear combination of the full constraints {\small{$\widetilde H[\lapse]$}} and
    {\small{$D[\shift]:=D_g[\shift]+D_m[\shift]$, $D_m[\shift]$}} being the matter contribution to the diffeomorphism constraint.
    It is easy to check that, when the modified matter Hamiltonian {\small{$\widetilde H_m[\lapse]$}} depends only on matter
    variables (including their radial derivatives) and triad components, but it does not depend on curvature components nor radial derivatives of the triad,
    the sum of the cross brackets,
        \begin{equation}
       \{\widetilde H_g[\lapse_1],\widetilde H_m[\lapse_2]\}+\{\widetilde H_m[\lapse_1],\widetilde H_g[\lapse_2]\},
    \end{equation}
    is vanishing due to antisymmetry. As there are no derivatives of $\krad$ or $\kang$ in the gravitational Hamiltonian and we are assuming no derivatives of the triad
    components in $H_m$, one does not need to perform any integration by parts when computing
    these Poisson brackets and, thus, both are proportional to the product $N_1 N_2$.
    
    Now, since the bracket between two vacuum Hamiltonians is proportional to the vacuum
    diffeomorphism constraint \eqref{eq:polhhgrav}, anomaly-freedom 
    demands that the bracket between matter
    Hamiltonians is proportional to the matter diffeomorphism constraint. In such
    a case, \eqref{fullhambracket} would read as follows,
    \begin{equation}
      \big\{\widetilde H_g[\lapse_1]+\widetilde H_m[\lapse_1],\widetilde H_g[\lapse_2]+\widetilde H_m[\lapse_2]\big\}=\int dr\,q^{rr} (\lapse_1\lapse_2'-\lapse_1'\lapse_2)\left[    \eta_1^2\!\frac{\partial\holbar}{\partial\kang} \diff_g+\Delta_m\diff_m\right],
    \end{equation}
    where, because of our assumptions,
    $\Delta_m$ depends on triad components and matter variables, but
    not on curvature components $\krad$ and $\kang$.
    In order to obtain a closed algebra,
    $\Delta_m$ should be equal to $\eta_1^2\,\partial\holbar\!/\partial\kang$,
    imposing that $\holbar$ does not depend on $\kang$ and thus ruling out
    the holonomy corrections that were allowed in vacuum.
    
    This general argument applies to simple matter models, such as minimally
    coupled scalar fields or dust. Nonetheless, one can try to bypass it by considering more generic relations. In fact,
    it might well happen that matter and geometric degrees of freedom develop non-minimal couplings as one approaches the quantum regime. In addition, one could expect polymerization, usually carried out only for the geometric degrees of freedom,
    to affect some of the matter variables, as proposed for instance in \cite{scalar1,scalar2,polscalar}.
       
    The object of this section is to study two different matter models, scalar field and dust,
    with a quite generic form of the deformed Hamiltonian. In particular, we will allow for non-minimal couplings
    as well as possible polymerization of both, geometric and matter degrees of freedom. More precisely, we will be assuming a general kinetic term for the modified Hamiltonian constraint, just as in the previous section. This generic term
    will be allowed to depend on any variable of the model and, thus, any coupling between
    the different matter and geometric variables will be possible. Furthermore, this general function might
    also take into account possible polymerization processes for all the degrees of freedom: not only for the curvature
    terms $\krad$ and $\kang$, but also for matter variables. In addition, multiplicative corrections depending
    on triad components and the corresponding matter field will be included on each of the remaining terms of the Hamiltonian, describing in that way possible inverse-triad corrections.
    At the same time, these functions will enable a more general coupling between matter and geometry.

    \subsection{Scalar field}\label{sec:scalarfield}  
    
  Assuming spherical symmetry, the contribution to the Hamiltonian and diffeomorphism constraints of a minimally coupled 
 scalar field, $\phi$,
are respectively given by,
\begin{align}
 {\cal H}_\phi&=\frac{\pphi\!^2}{2\sqrt{\erad}\ephi} 
      +\frac{(\erad)^{3/2}}{2\ephi}(\phi')^2+\sqrt{\erad}\ephi V(\phi)\:,\\
 {\cal D}_\phi&=\phi'\pphi \:.
\end{align}
The momentum $P_\phi$ is canonically conjugate to $\phi$, with the Poisson bracket,
\begin{equation}
 \{\phi(x), P_\phi(y)\}=\delta(x-y),
\end{equation}
that completes the phase-space structure given in \eqref{eq:vars}.

As in Section~\ref{sec:polgrav}, we will leave the diffeomorphism constraint unmodified,
${\cal D}={\cal D}_g+{\cal D}_\phi$
and will introduce the following initial proposal for the deformed Hamiltonian
constraint:
   \begin{align}
    \widetilde{\ham} &= f(\erad,\ephi,\phi,\krad,\kang,\pphi)+\eta_1(\erad,\ephi,\phi)\sqrt{\erad}{{\Gamma_{\!\varphi}}}\!' -\eta_2(\erad,\ephi,\phi)\frac{\ephi}{2\sqrt{\erad}}\nonumber\\&+\eta_3(\erad,\ephi,\phi)\frac{\ephi{\Gamma_{\!\varphi}\!}^2}{2\sqrt{\erad}}\label{modifiedhamiltonianphi}+\eta_4(\erad,\ephi,\phi)\frac{(\erad)^{3/2}}{2\ephi}(\phi')^2\:.
  \end{align}
The generic function $f$ of the deformed Hamiltonian is allowed to depend on all the canonical variables but not on their radial derivatives. It accounts for generic couplings
between different variables, as well as
inverse-triad and holonomy corrections (in particular, it encodes a possible polymerization of the matter variables). In addition, the free multiplicative functions that appear in the remaining terms stand for possible inverse-triad corrections and allow the scalar field to couple to radial derivatives of the triad.
   
   If one computes the Poisson brackets of the hypersurface deformation algebra with
   the above modified Hamiltonian,
   two anomalies are found, similar to those of the vacuum case (\ref{eq:anomalgdh}--\ref{eq:anomalghh}), but with a few additional matter terms. In order to get rid of these anomalous terms, one must demand them to be vanishing:
  \begin{align}\label{eq:anomaldh}
    0&=-4(\ephi)^2\eta_1\frac{\partial(\eta_2/\eta_1)}{\partial\ephi}+8\sqrt{\erad}\ephi\eta_1\frac{\partial(f/\eta_1)}{\partial\ephi}- 8\sqrt{\erad}\left(f-\krad\frac{\partial f}{\partial\krad}-\pphi\frac{\partial f}{\partial\pphi}\right) \nonumber\\
    &+\!\big[(\erad)'\big]^2\eta_1\frac{\partial(\eta_3/\eta_1)}{\partial\ephi}+4(\erad)^2\,(\phi')^2\eta_1\frac{\partial(\eta_4/\eta_1)}{\partial\ephi},\\\nonumber   \label{eq:anomalhh}
   0&=\, -\erad\ephi{\eta_1}\!^{2}(\ephi)'\frac{\partial^2(f/ \eta_1)}{\partial\ephi\partial\krad}-  \erad\ephi \eta_1\bigg[(\krad)'\frac{\partial^2f}{\partial\krad^2}+(\pphi)'\frac{\partial^2 f}{\partial\krad\partial\pphi}\bigg]\nonumber\\    &+(\erad)'\Bigg[\left(\eta_1-\eta_3\right)\frac{\ephi}{2}\frac{\partial f}{\partial\krad}-\erad\ephi{\eta_1}\!^{2}\frac{\partial^2(f/ \eta_1)}{\partial\erad\partial\krad}+\eta_1\erad\left(\frac{\partial f}{\partial\kang}-\krad\frac{\partial^2 f}{\partial\krad\partial\kang}\right)\! \Bigg] \nonumber\\
    &+\phi'\erad\ephi\Bigg[\,\eta_1\left(\frac{\pphi}{\ephi} \frac{\partial^2 f}{\partial\krad\partial\kang}- \frac{\partial^2 f}{\partial\krad\partial\phi}\right)+2\eta_4\erad\frac{\partial f}{\partial\pphi}+\frac{\partial\eta_1}{\partial\phi}\frac{\partial f}{\partial\krad}\Bigg]\:.
  \end{align}
  Since none of the free functions depends on derivatives of the different variables, the coefficient of each radial derivative in the above anomalies must be vanishing. Therefore, the last two conditions can be rewritten as a set of eight differential equations. In particular, the vanishing of the coefficients of $[(\erad)']^2$ and $(\phi')^2$ in \eqref{eq:anomaldh} is solved by the requirements,
  \begin{align}\label{eta3form}
    \eta_3(\erad,\ephi,\phi)&=\xi_3(\erad)\,\eta_1(\erad,\ephi,\phi)/\xi_1(\erad)\:,\\\label{eta4form}
    \eta_4(\erad,\ephi,\phi)&=\xi_4(\erad)\,\eta_1(\erad,\ephi,\phi)/\xi_1(\erad)\:,
  \end{align}
where $\xi_1$, $\xi_3$ and $\xi_4$ are free functions of $\erad$. In addition, as in the vacuum case, we have the freedom to choose $\eta_2$, as it could have been absorbed inside the generic function $f$ in definition \eqref{modifiedhamiltonianphi}. For convenience, we take,
\begin{equation}\label{eta2form}
  \eta_2(\erad,\ephi,\phi)=\xi_2(\erad)\,\eta_1(\erad,\ephi,\phi)/\xi_1(\erad)\:,
\end{equation}
which makes the first term in equation \eqref{eq:anomaldh} to vanish.

Considering now the restrictions for $f$, it is easy to see that the function
\begin{equation}\label{fform}
f=\frac{\eta_1(\erad,\ephi,\phi)}{\xi_1(\erad)} \Big[f_0(\erad,\ephi,\phi,\kang,\pphi)+\krad f_1(\erad,\phi,\kang)\Big] ,
\end{equation}
 with the integration functions $f_0$ and $f_1$ provides a generic solution to 
 the vanishing of the coefficients of $(\krad)'$, $(P_\phi)'$ and $(\ephi)'$ in \eqref{eq:anomalhh}. Thus, the dependence of the modified
Hamiltonian on the radial curvature component $\krad$ is bound to be linear and, therefore, holonomy corrections for this variable will not be allowed. Replacing now the results \eqref{eta3form}--\eqref{fform} in equation \eqref{eq:anomaldh}, we obtain the following condition for $f_0$:
\begin{equation}
 f_0-\pphi \frac{\partial f_0}{\partial \pphi}-\ephi \frac{\partial f_0}{\partial \ephi}=0\:,
\end{equation}
which is the generalization of the vacuum relation \eqref{f0f1relation}.
This equation restricts the form of the function $f_0$
in terms of the variables $\ephi$ and $P_\phi$. More precisely, $f_0$ must be of the form $f_0=\ephi f_2 (\erad, \phi, \kang, \pphi/\ephi)$, with $f_2$ an integration function. Therefore, we can rewrite $f$ above, as
\begin{equation}\label{fform2}
f=\frac{\eta_1(\erad,\ephi,\phi)}{\xi_1(\erad)}\Big[\ephi f_2 (\erad, \phi, \kang, \pphi/\ephi)+\krad f_1(\erad,\phi,\kang)\Big]\:. 
\end{equation}
Expressions \eqref{eta3form}--\eqref{eta2form} and \eqref{fform2} completely solve equation \eqref{eq:anomaldh}. Replacing all the results in the other anomaly \eqref{eq:anomalhh}, every term vanishes except for the coefficients of $(\erad)'$ and $\phi'$, which provide the following two additional restrictions:
  \begin{align}\label{eq:anodiff1}
    \frac{\partial f_2}{\partial\kang} &= \frac{\xi_3-\xi_1}{2\erad\xi_1}f_1+\xi_1\frac{\partial (f_1/\xi_1)}{\partial\erad}, \\ 
 \label{eq:anodiff2}
    \frac{\partial f_2}{\partial(\pphi/\ephi)} &=\frac{\xi_1}{2\erad\xi_4}\left(  \frac{\partial f_1}{\partial\phi} - \frac{\pphi}{\ephi}\frac{\partial f_1}{\partial\kang}\right).
  \end{align}
Note that equation \eqref{eq:anodiff1} is the equivalent to
 \eqref{eq:anomalg} for this model (in order to see this, it is
 enough to compare the decomposition \eqref{fform2}
 with \eqref{eq:fgfin}). But in this case, besides \eqref{eq:anodiff1}, the functions $f_1$ and $f_2$ also need to satisfy the relation \eqref{eq:anodiff2}. This will completely fix the dependence of these functions on the angular component of the curvature $\kang$, preventing in this way the presence of free functions of this variable. Therefore, the fundamental reason for this model not to admit holonomy corrections can be pinpointed to the existence of this last equation.

  In order to solve the last two equations, let us first focus on the dependence of the functions $f_1$ and $f_2$ on the variables $\kang$ and $P_\phi/\ephi$. Since the right-hand side of equation \eqref{eq:anodiff1} does not contain any $P_\phi/\ephi$ terms, these variables cannot be coupled to $\kang$. That is, $f_2$ must be a sum of two functions: $f_2=g_1(\erad,\phi,P_\phi/\ephi)+g_2(\erad,\phi,\kang)$. Plugging this form in \eqref{eq:anodiff2} and taking the derivative of that equation with respect to $\kang$, one obtains:
  \begin{equation}
   \frac{\partial^2 f_1}{\partial\kang\partial\phi} - \frac{\pphi}{\ephi}\frac{\partial^2 f_1}{\partial\kang^2}=0\:.
  \end{equation}
  It is clear that, since $f_1$ does not depend on $\pphi/\ephi$, each term of this
  equation must vanish independently. Then, one concludes that $f_1$ is, at most, linear
  in $\kang$, with the coefficient of $\kang$ being independent of the scalar field variable $\phi$:
  \begin{align}\label{f3form}
   f_1= \xi_1(\erad)\Big[f_{\phi}(\erad,\phi) +\kang f_{k}(\erad)\Big]\:.
  \end{align}
  As already commented above, this result provides the explicit dependence of the function $f_1$ on
  the variables $\kang$ and $\pphi/\ephi$. Therefore, one can replace
  \eqref{f3form} in the system \eqref{eq:anodiff1}--\eqref{eq:anodiff2}
  and integrate the equations with respect to the mentioned variables.
  The explicit dependence of $f_2$ on $\kang$ and $P_\phi/\ephi$ reads as follows,
  \begin{align}\label{f2form}
  f_2&= \xi_1(\erad)\Big[f_{00}(\erad,\phi) +\kang f_{10}(\erad,\phi) +\kang^2 f_{20}(\erad,\phi)\nonumber\\&+\frac{\pphi}{\ephi}\,f_{01}(\erad,\phi) +\bigg(\frac{\pphi}{\ephi}\bigg)^{\!\!2}\! f_{02}(\erad,\phi)\Big].
  \end{align}
  Replacing the above expressions for $f_1$ \eqref{f3form} and $f_2$ \eqref{f2form} in equations \eqref{eq:anodiff1}-\eqref{eq:anodiff2}, we obtain a set of four differential relations, one for each coefficient of different powers in $\kang$ and $P_\phi/\ephi$:
  \begin{align}\label{eq:eqfinal1}
      f_{02}&=-\frac{\xi_1}{4\erad\xi_4} f_{k}\:,\\
      f_{20} &=  \frac{\xi_3-\xi_1}{4\erad\xi_1}f_{k}+\frac{1}{2}\frac{\partial f_{k}}{\partial\erad}\:,\label{eq:eqfinal2}\\ 
      f_{10} &=  \frac{\xi_3-\xi_1}{2\erad\xi_1}f_{\phi}+\frac{\partial f_{\phi}}{\partial\erad}\:,\label{eq:eqfinal3}\\
         f_{01}&=\frac{\xi_1}{2\erad\xi_4}\frac{\partial f_{\phi}}{\partial\phi}\label{eq:eqfinal4}\:.
  \end{align}
 In particular, since $f_k$ depends only on the radial component of the triad $\erad$, equations \eqref{eq:eqfinal1} and \eqref{eq:eqfinal2} show that neither $f_{02}$ nor $f_{20}$ depends on $\phi$. 
 Out of the seven free functions in \eqref{f3form}--\eqref{f2form}, only three will remain free after these last four equations are satisfied. The function $f_{00}$ is one of those free
 functions, as it does not appear in the last relations. In addition, the functions $f_k$ and $f_\phi$ will be chosen as free, whereas the remaining functions can be written in terms of these two.
 Finally, we rename two of the free functions with the definitions $\mathcal{V}(\erad,\phi):=\xi_1f_{00}/\sqrt{\erad}$ and $\xi_0:=-f_k/(2\sqrt{\erad})$ to resemble the classical form of the Hamiltonian. Taking into account all the above results, the modified Hamiltonian constraint is given by,
   \begin{align}\label{eq:polhamscalar}
            \widetilde{\ham} &= \eta(\erad,\ephi,\phi)      
      \Bigg[\sqrt{\erad}\xi_1\Big({\Gamma_{\!\varphi}}'-2\xi_0\krad\kang\Big) -\frac{\ephi}{2\sqrt{\erad}}\bigg(\xi_2+\left[{\xi_0\xi_3} +2\erad\xi_1\frac{\partial\xi_0}{\partial\erad}\right] \!{\kang}\!^2 -\xi_3{\Gamma_{\!\varphi}\!}^2\bigg)  \nonumber\\    &+\frac{\xi_0 {\xi_1}\!^2}{\xi_4}\frac{{\pphi}\!^2}{2\sqrt{\erad}\ephi}
      +  \xi_4\frac{(\erad)^{3/2}}{2\ephi}(\phi')^2
      +\sqrt{\erad}\ephi\mathcal{V}(\phi,\erad)+\xi_1\krad f_\phi+\frac{{\xi_1}\!^2\pphi}{2\erad\xi_4}\frac{\partial f_\phi}{\partial\phi}\nonumber\\ &+\bigg( \frac{\xi_3-\xi_1}{2\erad}f_{\phi}+\xi_1\frac{\partial f_{\phi}}{\partial\erad}\bigg)\ephi\kang\Bigg],     
    \end{align}
  where the free functions $\xi_i=\xi_i(\erad)$, for $i=0,1,2,3,4$, would be interpreted
  as inverse-triad corrections and depend only on the radial component of the triad, whereas
  the global factor $\eta:=\eta_1/\xi_1$ depends on all the configuration
  variables.
  In addition, the function ${\cal V}={\cal V}(\erad,\phi)$ stands for a generalized potential term
  for the scalar field, with an arbitrary dependence on the radial
  triad $\erad$. Finally, the last free function $f_\phi=f_\phi(\erad,\phi)$ appears
  in linear terms of the momenta $\krad$, $\kang$, and $P_\phi$ and
  couples the scalar field to curvature components. However, this function
  does not encode any physical information and
  can be removed by the following canonical transformation,
  \begin{align}\label{can1}
      \bar{K}_\varphi &= \kang -\frac{f_\phi}{2\sqrt{\erad}\xi_0}\:,\\\label{can2}
     \bar{K}_r &= \krad -\frac{\partial}{\partial\erad}\left(\frac{\ephi f_\phi}{2\sqrt{\erad}\xi_0}\right)\:,\\\label{can3}
      \bar{P}_\phi &= \pphi +\frac{\ephi}{2\sqrt{\erad}\xi_0}\frac{\partial f_\phi}{\partial\phi}\:.
  \end{align}
  Remarkably, this transformation leaves the diffeomorphism constraint invariant. In terms of these new variables, the final form of
  the modified Hamiltonian constraint reads as follows,
   \begin{align}\label{eq:polhamscalargauge}
     \widetilde{\ham} &= \eta(\erad,\ephi,\phi)     
     \Bigg[\sqrt{\erad}\xi_1\Big({\Gamma_{\!\varphi}}'-2\xi_0\bar{K}_r\bar{K}_\varphi\Big)-\frac{\ephi}{2\sqrt{\erad}}\bigg(\xi_2+\left[{\xi_0\xi_3} +2\erad\xi_1\frac{\partial\xi_0}{\partial\erad}\right] \!{\bar{K}_\varphi}\!^2 -\xi_3{\Gamma_{\!\varphi}\!}^2\bigg)  \nonumber\\    &+\frac{\xi_0 {\xi_1}\!^2}{\xi_4}\frac{\bar{P}_\phi\!^2}{2\sqrt{\erad}\ephi}
      +  \xi_4\frac{(\erad)^{3/2}}{2\ephi}(\phi')^2
      +\sqrt{\erad}\ephi\mathcal{\bar{V}} \Bigg],
          \end{align}
where the potential term ${\cal{V}}(\phi,\erad)$ has been redefined as a new generic function ${\cal\bar{V}}(\phi,\erad)$
in order to absorb the extra terms from the canonical transformation. The classical Hamiltonian is recovered when
all the correction functions, $\eta$ and $\xi_i$, are equal to one and $\bar{\mathcal{V}}$ stands for the scalar field potential $V(\phi)$.

Notice that the dependence of this Hamiltonian on the
momenta is the same as in the classical case. In particular, there are no free functions
of the curvature components and thus, in contrast to the vacuum case,
no holonomy corrections are allowed in this model.

Remarkably, even if we started from a generic
and coupled form \eqref{modifiedhamiltonianphi}, the final modified Hamiltonian constraint, obtained through the requirement of anomaly freedom, can be expressed as a direct sum
of a corrected vacuum and scalar-field Hamiltonian constraints, {\small$\widetilde{\ham}=\widetilde{\ham}_g^{(m)}+\widetilde{\ham}_{s}$}, which
read, respectively, as follows:
  \begin{align}\label{modhamg}
      \widetilde{\ham}_g^{(m)} &:= \eta    \Bigg[\sqrt{\erad}\xi_1\Big({\Gamma_{\!\varphi}}'-2\xi_0\bar{K}_r\bar{K}_\varphi\Big)-\frac{\ephi}{2\sqrt{\erad}}\bigg(\xi_2 \!+\!\bigg[{\xi_0\xi_3} \!+\!2\erad\xi_1\frac{\partial\xi_0}{\partial\erad}\bigg]\bar{K}_\varphi\!^2 -\xi_3{\Gamma_{\!\varphi}\!}^2\bigg)\! \Bigg],\\
      \widetilde{\ham}_{s} &:= \eta\Bigg[
      \frac{\xi_0{\xi_1}\!^2}{\xi_4}\frac{\bar{P}_\phi\!^2}{2\sqrt{\erad}\ephi} 
      +\xi_4\frac{(\erad)^{3/2}}{2\ephi}(\phi')^2+\sqrt{\erad}\ephi\mathcal{\bar{V}}(\phi,\erad)\Bigg]\:.
  \end{align}
 Therefore, the only non-minimal coupling between matter and geometric degrees of freedom
 is given by the global factor $\eta$. Note, however, that the modified Hamiltonian
 constraint corresponding to the geometric degrees of freedom \eqref{modhamg} is not the same as the one obtained in the vacuum model \eqref{eq:polhamgrav}. As already commented above,
 in this case we have less freedom than in the pure vacuum model:
 instead of two functions $h_1$ and $h_2$ of two variables $(\erad,\kang)$ related by equation \eqref{eq:anomalg},
 in this case we have just one free function $\xi_0$ of the radial component of the triad $\erad$.

  If we now compute the Poisson brackets using the modified Hamiltonian constraint \eqref{eq:polhamscalargauge},   the hypersurface deformation algebra takes the following form:
  \begin{align}\label{eq:poldh}
    \{ D[\shift] , \widetilde{H}[\lapse] \} &=\widetilde{H}\bigg[\shift\lapse'-(\shift)'\lapse\frac{\partial{\eta}}{\partial\ephi}\frac{\ephi}{\eta}\bigg],\\
\label{eq:polhh}
    \{ \widetilde{H}[\lapse_1] , \widetilde{H}[\lapse_2] \} &= D\Big[\eta^2\xi_0{\xi_1}\!^2\, q^{rr}(\lapse_1{\lapse_2}\!'-{\lapse_1}\!'\lapse_2)\Big].
  \end{align}
The free functions $\eta$, $\xi_0$ and $\xi_1$ in the last bracket are now the ones that might produce physical effects such as signature change.

\subsubsection*{Singular solutions}

 Concerning the solutions not included in the previous derivation, one can obtain a modified Hamiltonian
  that forms an anomaly-free algebra, along with the classical diffeomorphism constraint, 
  demanding one of the following conditions:
   \begin{changemargin}{0.5cm}{0cm}
  \begin{enumerate}[label=(\roman*)]
  \item $\eta_1=0$, $\eta_3=0$, and $f=f(\erad, \ephi,\phi,\krad,\kang)$.\label{cs1}
\item $\eta_1=0$, $\eta_4=0$, and $f=f(\erad, \ephi,\phi,\kang,P_\phi)$.\label{cs2}
\item $\eta_1=0$, and $f=f(\erad, \ephi,\phi,\kang)$.\label{cs3}
   \item $\eta_1=0$, $\eta_3=0$, and $\eta_4=0$.\label{cs4}
\end{enumerate}
\end{changemargin}

Taking $\eta_1=0$, one removes the radial derivative of the spin-connection
and, thus, terms containing $(\erad)''$ and $(\ephi)'$
are pulled out from the Hamiltonian. In addition, two
further conditions are required. For example, one can demand $\eta_3=0$, removing
all radial derivatives of the triad components and that the Hamiltonian
does not depend on the scalar field momentum, that is, case \ref{cs1}. Alternatively, one can take $\eta_4=0$, which would remove the radial derivatives of the scalar field, and request that the Hamiltonian does not depend on $\krad$ (case \ref{cs2}). Another possibility, \ref{cs3}, implies the removal of the momenta $\krad$ and $\pphi$ from the Hamiltonian, leaving $\eta_3$ and $\eta_4$ free. The last option would be to eliminate all radial derivatives (case \ref{cs4}). As in the pure vacuum model, this last case resembles the corresponding homogeneous spacetime.

In summary, holonomy corrections for a given momentum ($\krad$, $\kang$, $P_\phi$)
are incompatible with the presence of the radial derivative of its corresponding
conjugate variable ($(\erad)'$, $(\ephi)'$, $\phi'$) in the modified Hamiltonian. Nevertheless, taking into account the BKL conjecture,
one could consider a model where, as small values of the triad components were approached,
certain correction functions would be vanishing from a given value on, annihilating
in this way radial derivatives in the Hamiltonian and entering the homogeneous BKL limit.
In this homogeneous region, holonomy corrections
would be allowed with no further restriction arising from anomaly freedom. Such a model would be described, for instance, by a modified Hamiltonian of the form,
\begin{align}
    \widetilde{\ham} &= \frac{\widetilde f_1(P_\phi)}{2\sqrt{\erad}\ephi}
     -2\sqrt{\erad} \widetilde f_2(\krad, \kang)  -\frac{\ephi}{2\sqrt{\erad}} \widetilde f_3(\kang)
    \label{hamiltonianwithholonomies}\nonumber\\&+\eta_1\sqrt{\erad}{\Gamma_{\!\varphi}}'
    +\eta_3\frac{\ephi{\Gamma_{\!\varphi}\!}^2}{2\sqrt{\erad}}+\eta_4\frac{(\erad)^{3/2}}{2\ephi}(\phi')^2\:.
  \end{align}
  This Hamiltonian would form a first-class algebra with the classical diffeomorphism constraint if
  ${\widetilde f_1}$, ${\widetilde f_2}$, and ${\widetilde f_3}$ took their corresponding classical value
  as long as $\eta_1$, $\eta_3$, and $\eta_4$ were non-vanishing. In the region of the configuration space
  where $\eta_1=\eta_3=\eta_4=0$, the new functions ${\widetilde f_1}$, ${\widetilde f_1}$, and ${\widetilde f_3}$
  would be completely free and could encode, for instance, holonomy modifications.

  \subsection{Dust}\label{sec:dust}
  
    In spherical symmetry, a dust field can be described in terms of a scalar variable $\Phi$ and
  its conjugate momentum $ P_\Phi$, with Poisson brackets $\{\Phi(x), P_\Phi(y)\}=\delta(x-y)$. The classical contributions of the dust
  to the Hamiltonian and diffeomorphism constraints are respectively given by,
  \begin{align}
   {\cal H}_\Phi&=  P_\Phi\sqrt{1+\frac{\erad(\Phi')^2}{(\ephi)^2}}\:,\\
   {\cal D}_\Phi&= \Phi' P_\Phi\:.
  \end{align}
Even if, due to the presence of the square root in the dust Hamiltonian constraint, the computations will be a bit more complicated than in the previous case,
the procedure will be similar. As the starting point of our analysis, we will consider the following general
form for the modified Hamiltonian that describes the dynamics of the dust model
coupled to gravity,
   \begin{align}\label{genericdusthamiltonian}
    \widetilde{\ham} &=f(\erad,\ephi,\Phi,\krad,\kang, P_\Phi)
    +\eta_1(\erad,\ephi,\Phi)\sqrt{\erad}{\Gamma_{\!\varphi}}\!'-\eta_2(\erad,\ephi,\Phi)\frac{\ephi}{2\sqrt{\erad}}\nonumber\\&+\eta_3(\erad,\ephi,\Phi)\frac{\ephi{\Gamma_{\!\varphi}\!}^2}{2\sqrt{\erad}}+g(\erad,\ephi,\Phi,\krad,\kang, P_\Phi)\sqrt{1+\eta_4(\erad,\ephi,\Phi)\frac{\erad(\Phi')^2}{(\ephi)^2}}
  \end{align}
and let us define the shorthand notation 
\begin{equation}
  \mathcal{S}:=\sqrt{1+\eta_4(\erad,\ephi,\Phi)\frac{\erad(\Phi')^2}{(\ephi)^2}},
\end{equation}  
has been defined to encode the square root.

Notice that we have introduced two generic functions, $f$ and $g$, that might
depend on all the variables of the model, but not on their radial derivatives.
These functions replace the classical terms that go with the momenta $\krad$,
$\kang$, and $P_\Phi$. More precisely, the function $f$ replaces the quadratic
combination of curvature components in the geometric Hamiltonian, whereas
$g$ takes the place of the dust field momentum $P_\Phi$. In this way,
these two functions might describe the polymerization of the momenta
but also, as they depend on the configuration variables, might stand for inverse-triad
corrections. In addition, their generic form is suitable for any kind of
coupling between geometric and matter degrees of freedom. Finally, the functions
$\eta_i$, with $i=1,2,3,4$, are allowed to depend on the triad components
$\erad$ and $\ephi$ in order to stand for possible inverse-triad corrections. Additionally, they might also depend on the dust field, enabling a generic coupling.

As in the previous models, two anomalies are obtained when the
Poisson algebra is computed. In order to obtain a first class algebra, we demand that both anomalous terms vanish:

    \begin{align}\label{eq:anomaldhdust}
    0 &= 
    2\eta_1(\ephi)^2 \mathcal{S}\frac{\partial (\eta_2/\eta_1)}{\partial\ephi}     -4\eta_1\sqrt{\erad}\ephi \mathcal{S}\frac{\partial}{\partial\ephi}\Bigg[\frac{f +g \mathcal{S}}{\eta_1}\Bigg]+g\eta_1\eta_4(\Phi')^2\frac{(\erad)^{3/2}}{(\ephi)^2}\nonumber\\  &+4\sqrt{\erad} \mathcal{S} \left(1- P_\Phi\frac{\partial}{\partial P_\Phi}-\krad\frac{\partial}{\krad}\right)\!\!\big(f+ \mathcal{S}g\big)    -\eta_1\big[(\erad)'\big]^2\mathcal{S}\frac{\partial (\eta_3/\eta_1)}{\partial\ephi} \:,\\
  \label{eq:anomalhhdust}
    0 &= \eta_1\eta_4\erad(\ephi)^{-1}\Phi'\Phi''\frac{\partial g}{\partial\krad}
  \eta_1\ephi \mathcal{S}\left((\krad)'\frac{\partial}{\partial\krad}+( P_\Phi)'\frac{\partial}{\partial P_\Phi}\right)\!\!\left(\frac{\partial f}{\partial\krad} + \mathcal{S}\frac{\partial g}{\partial\krad}\right)\nonumber\\    &-\eta_1(\erad)'\mathcal{S}\Bigg[\!\!\left(1\!-\!\krad\frac{\partial}{\partial\krad}\!\right)\!\!\left(\frac{\partial f}{\partial\kang}\!+\!\mathcal{S}\frac{\partial g}{\partial\kang}\right) \! +\!\ephi\!\left(\frac{\eta_1-\eta_3}{2\erad}\!-\!\frac{\partial}{\partial\erad}\!\right)\!\!\left(\frac{\partial (f/\eta_1)}{\partial\krad}\!+\!\mathcal{S}\frac{\partial (g/\eta_1)}{\partial\krad}\right) \!\! \Bigg]\nonumber\\    &-\Phi'\mathcal{S}\Bigg[\eta_1\left(\! P_\Phi\frac{\partial }{\partial\kang}\!-\ephi\frac{\partial }{\partial\Phi}\right)\!\!\left(\frac{\partial f}{\partial\krad}+\mathcal{S}\frac{\partial g}{\partial\krad}\right)+\ephi\frac{\partial\eta_1}{\partial\Phi}\left(\frac{\partial f}{\partial\krad}+\mathcal{S}\frac{\partial g}{\partial\krad}\right)  \nonumber\\&+2\sqrt{\erad}\eta_4 g\left(\mathcal{S}^{-1}\frac{\partial f}{\partial P_\Phi} +\frac{\partial g}{\partial P_\Phi}\right)\!\Bigg]+{\eta_1}\!^2\ephi(\ephi)' \mathcal{S}\Bigg[\frac{\partial }{\partial\ephi}\!\left(\frac{\partial (f/\eta_1)}{\partial\krad}+ \mathcal{S}\frac{\partial( g/\eta_1)}{\partial\krad}\right)\Bigg].
  \end{align}
  Due to the presence of the square root in ${\cal S}$ and the coupling between $ P_\Phi$ and $\Phi'$, these anomalies are a bit more involved than those in Sec.~\ref{sec:scalarfield}. Nevertheless, some conditions for the free functions can immediately be read out. In particular, from the vanishing of the coefficient of $[(\erad)']^2$ in equation \eqref{eq:anomaldhdust}, one obtains:
  \begin{equation}\label{eta3formdust}
  \eta_3(\erad,\ephi,\Phi)=\xi_3(\erad)\,\eta_1(\erad,\ephi,\Phi)/\xi_1(\erad)\:.
  \end{equation}
  In addition, as in previous cases, we use the freedom in the definition of the function $\eta_2$ to choose,
  \begin{equation}\label{eta2formdust}
  \eta_2(\erad,\ephi,\Phi)=\xi_2(\erad)\,\eta_1(\erad,\ephi,\Phi)/\xi_1(\erad)\:,
  \end{equation}
  which annihilates the first term of the anomaly \eqref{eq:anomaldhdust}. The first term in \eqref{eq:anomalhhdust} is the only one with
  second-order radial derivatives of the dust field and must vanish on its own. This demand is satisfied when the function $g$ does not depend on $\krad$. Therefore, we will write this function as,
  \begin{equation}\label{gformdust}
  g=\eta_1(\erad,\ephi,\phi)\,g_0(\erad,\ephi,\kang,\Phi, P_\Phi)/\xi_1(\erad)\:,
  \end{equation}
  where the function $\eta_1/\xi_1$ has been introduced as a global factor for convenience.  
  As there is no more dependence on radial derivatives of the variables, the coefficients of $(\krad)'$, $( P_\Phi)'$, and $(\ephi)'$ in the anomaly \eqref{eq:anomalhhdust} must also vanish by themselves. Taking into account \eqref{gformdust}, three independent equations for the function $f$ are obtained:
  \begin{align}
   0&=\frac{\partial^2 f}{\partial K_r^2}\:,\\
   0&=\frac{\partial^2 f}{\partial K_r\partial  P_\Phi}\:,\\
   0&=\frac{\partial^2 (f/\eta_1)}{\partial\ephi\partial K_r}\:.
  \end{align}
  It is straightforward to see that the general solution for this system can be written in terms of two free functions, $f_0$ and $f_1$, as follows,
 \begin{align}\label{hformdust}
  f=\frac{\eta_1}{\xi_1}\big[f_0(\erad,\ephi,\Phi,\kang, P_\Phi)+\krad f_1(\erad,\Phi,\kang)\big].
 \end{align}
At this point, we already see that the modified Hamiltonian will be linear in $\krad$ and that this variable will be decoupled from the momentum of the dust field $ P_\Phi$.
 
The replacement of \eqref{eta3formdust}--\eqref{gformdust} and \eqref{hformdust} in the anomaly equations \eqref{eq:anomaldhdust}--\eqref{eq:anomalhhdust} leads to a much simpler version of these two equations:
 \begin{align}\label{eq:anomaldhdust2}
    0 &=  
    \mathcal{S}\bigg(f_0-\ephi\frac{\partial f_0}{\partial\ephi}- P_\Phi\frac{\partial f_0}{\partial P_\Phi}\bigg)+\bigg(g_0-\ephi\frac{\partial g_0}{\partial\ephi}- P_\Phi\frac{\partial g_0}{\partial P_\Phi}\bigg)\nonumber\\ &+\frac{\eta_4\erad}{(\ephi)^2}(\Phi')^2\,\bigg(-\frac{\ephi}{2}\frac{g_0}{\eta_4} \frac{\partial\eta_4}{\partial\ephi}+g_0-\ephi\frac{\partial g_0}{\partial\ephi}- P_\Phi\frac{\partial g_0}{\partial P_\Phi}\bigg)\:,\\\nonumber   0 &=
     \Phi'\Bigg[ P_\Phi\frac{\partial f_1}{\partial\kang}-\ephi\frac{\partial f_1}{\partial\Phi}+2\sqrt{\erad}\eta_4 \frac{g_0}{\xi_1}\left( \mathcal{S}^{-1}\frac{\partial f_0}{\partial P_\Phi}+\frac{\partial g_0}{\partial P_\Phi}\right)\!\Bigg]\nonumber\\ &+(\erad)'\Bigg[\ephi\left(\frac{\xi_1-\xi_3}{2\erad\xi_1}f_1-\xi_1\frac{\partial (f_1/\xi_1)}{\partial\erad}\right)+\frac{\partial f_0}{\partial\kang}+\mathcal{S}\frac{\partial g_0}{\partial\kang} \Bigg]\:.\label{eq:anomalhhdust2} 
  \end{align}
  In fact, looking at the dependence on the radial derivatives of the variables, these two equations can be rewritten as a set of seven independent equations. From \eqref{eq:anomaldhdust2}, one gets the following three differential equations,
  \begin{align}\label{e1}
    0&=g_0-\ephi\frac{\partial g_0}{\partial\ephi}- P_\Phi\frac{\partial g_0}{\partial P_\Phi}\:,\\\label{e2}
    0&= f_0-\ephi\frac{\partial f_0}{\partial\ephi}- P_\Phi\frac{\partial f_0}{\partial P_\Phi}\:,\\\label{e3}
    0&=\frac{\ephi}{2}\frac{g_0}{\eta_4}\frac{\partial\eta_4}{\partial\ephi}
        -\left(g_0-\ephi\frac{\partial g_0}{\partial\ephi}- P_\Phi\frac{\partial g_0}{\partial P_\Phi}\right),
  \end{align}
  whereas \eqref{eq:anomalhhdust2} provides four additional conditions:
  \begin{align}\label{e5}
    0&=\frac{\partial f_0}{\partial  P_\Phi}\:,\\\label{e6}
    0&=\frac{\partial g_0}{\partial\kang}\:,\\
    \label{e4}
   0&= P_\Phi\frac{\partial f_1}{\partial\kang}-\ephi\frac{\partial f_1}{\partial\Phi}
        +2\sqrt{\erad}\eta_4 \frac{g_0}{\xi_1}\frac{\partial g_0}{\partial P_\Phi}\:,\;\;\\\label{e7}
    0&=\ephi\left(\frac{\xi_1-\xi_3}{2\erad\xi_1}f_1-\xi_1\frac{\partial (f_1/\xi_1)}{\partial\erad}\right)
        +\frac{\partial f_0}{\partial\kang}\:.
  \end{align}
Equation \eqref{e5} demands $f_0$ to be independent of $ P_\Phi$ and
 one deduces that $g_0$ cannot depend on $\kang$ from equation \eqref{e6}. Hence, there will not be holonomy modifications multiplying the $\Phi'$ term of the Hamiltonian. Using these results, the general solutions to \eqref{e1} and \eqref{e2} are respectively given by:
  \begin{align}
   g_0&=\ephi\,\widetilde{g_0}\left(\erad,\Phi,\frac{ P_\Phi}{\ephi}\right)\:,\\
   f_0&=\ephi\,\widetilde{f}_0\left(\erad,\kang,\Phi\right)\:.
  \end{align}  
Note that the term between brackets in equation \eqref{e3}
is equal to the right-hand side of equation \eqref{e1}, so it must vanish. Therefore, it is straightforward to integrate \eqref{e3}
to obtain that $\eta_4=1/\xi_4(\erad,\Phi)$, with the free integration
function $\xi_4$.

  From the system
  \eqref{e1}--\eqref{e7}, only two differential equations
  are left to be solved, namely \eqref{e4} and \eqref{e7}.
  Enforcing the solutions for the other equations, relations
  \eqref{e4} and \eqref{e7} are reduced to the following form:
  \begin{align}
  \label{h0h1}
      \frac{\partial \widetilde{f}_0}{\partial\kang}&=\frac{\xi_3-\xi_1}{2\erad\xi_1}f_1+\xi_1\frac{\partial (f_1/\xi_1)}{\partial\erad}\:,\\\label{g0tilde}
        \frac{\partial\big( \widetilde{g_0}^2\big)}{\partial( P_\Phi/\ephi)} &= \frac{\xi_1\xi_4}{\sqrt{\erad}}\left(\frac{\partial f_1}{\partial\Phi}-\frac{ P_\Phi}{\ephi}\frac{\partial f_1}{\partial\kang}\right)\:.
  \end{align}
  If we differentiate the first one with respect to $\kang$, two independent conditions arise:
   \begin{align}
     0&=\frac{\partial^2 f_1}{\partial \kang^2}\:,\\
     0&=\frac{\partial^2 f_1}{\partial \kang\partial \Phi}\:,
   \end{align}
   which are solved by the expression,
  \begin{equation}\label{h1}
      f_1= \xi_1(\erad)\Big[f_\Phi(\erad,\Phi)-2\sqrt{\erad}\xi_0(\erad)\kang\Big]\:.
  \end{equation}
   The integration functions $\xi_0$ and $f_\Phi$
  have been chosen to resemble the notation of previous sections.
  Plugging this form for $f_1$ in \eqref{h0h1} and \eqref{g0tilde}, it is straightforward to integrate those two equations
  and obtain the general form of the functions $\widetilde g_0$
  and $\widetilde f_0$:
  \begin{align}
     \!\!\!\! \widetilde{g_0} &= \xi_1\sqrt{\xi_4}\sqrt{\xi_0\left(\frac{ P_\Phi}{\ephi}\right)^2 +\frac{ P_\Phi}{\sqrt{\erad}\ephi}\frac{\partial  f_\Phi}{\partial\Phi}+\mathcal{V}_1(\erad,\Phi)}\:,\\
      \widetilde{f}_0 &= -\bigg(\frac{\xi_0\xi_3}{2\sqrt{\erad}}+\sqrt{\erad}\xi_1\frac{\partial \xi_0}{\partial\erad}\bigg)\kang^2+\left(\frac{\xi_3-\xi_1}{2\erad} f_\Phi+\xi_1\frac{\partial  f_\Phi}{\partial\erad}\right)\kang +\mathcal{V}_2(\erad,\Phi)\:.
  \end{align}
  The terms $\mathcal{V}_1(\erad,\Phi)$
  and $\mathcal{V}_2(\erad,\Phi)$ are completely free integration functions.

  Finally, collecting all the results, one arrives to the following modified Hamiltonian constraint:
   \begin{align}
     \widetilde{\ham} &= \eta(\erad,\ephi,\Phi)    
      \Bigg[\sqrt{\erad}\xi_1\Big({\Gamma_{\!\varphi}}'-2\xi_0\krad\kang\Big) -\frac{\ephi}{2\sqrt{\erad}}\bigg(\xi_2 +\bigg[\xi_0\xi_3 +2\erad\xi_1\frac{\partial\xi_0}{\partial\erad}\bigg]{\kang}\!^2 -\xi_3{\Gamma_{\!\varphi}\!}^2\bigg) \nonumber\\  &+\xi_1\sqrt{\xi_0 P_\Phi^2+\frac{\ephi P_\Phi}{\sqrt{\erad}}\frac{\partial  f_\Phi}{\partial\Phi}+(\ephi)^2\mathcal{V}_1(\erad,\Phi)\,}\,\sqrt{\xi_4+\frac{\erad(\Phi')^2}{(\ephi)^2}} +\sqrt{\erad}\ephi\mathcal{V}_2(\erad,\Phi)+\xi_1\krad  f_\Phi\nonumber\\&+\left(\frac{\xi_3-\xi_1}{2\erad} f_\Phi+\xi_1\frac{\partial  f_\Phi}{\partial\erad}\right)\ephi\kang\, \Bigg]\:,
    \end{align}
  with the global factor defined as $\eta:=\eta_1/\xi_1$.
Note that $f_\Phi$ plays a similar role as the function $f_\phi$
introduced in the previous section, and appears multiplying linear terms
of the momenta. Indeed, as happened with $f_\phi$, one can perform a canonical
transformation that removes it from the Hamiltonian and preserves the form of the diffeomorphism constraint:
   \begin{align}
     \bar K_\varphi&= K_\varphi-\frac{ f_\Phi}{2\xi_0\sqrt{\erad}}\:,\\
     \bar K_r&= K_r-\frac{\partial}{\partial\erad}\left(\frac{\ephi f_\Phi}{2\sqrt{\erad}\xi_0}\right)\,,\\
     \bar  P_\Phi&=  P_\Phi+\frac{\ephi}{2\xi_0\sqrt{\erad}}\frac{\partial  f_\Phi}{\partial\Phi}\:.
    \end{align}
     Considering these new canonical variables, the modified Hamiltonian constraint reads:
   \begin{align}\label{eq:polhamdust}
     \widetilde{\ham} &= \eta(\erad,\ephi,\Phi)  
      \Bigg[\sqrt{\erad}\xi_1\Big({\Gamma_{\!\varphi}}'-2\xi_0\bar\krad\bar{K}_\varphi\Big)-\frac{\ephi}{2\sqrt{\erad}}\bigg(\xi_2 +\bigg[\xi_0\xi_3 +2\erad\xi_1\frac{\partial\xi_0}{\partial\erad}\bigg]{\bar K}_\varphi^2 -\xi_3{\Gamma_{\!\varphi}\!}^2\bigg) \nonumber\\
      &+\xi_1\sqrt{\xi_0 P_\Phi^2+(\ephi)^2\overline{\mathcal{V}}_1(\erad,\Phi)}\sqrt{\xi_4+\frac{\erad(\Phi')^2}{(\ephi)^2}} +\sqrt{\erad}\ephi\overline{\mathcal{V}}_2(\erad,\Phi) \Bigg].
    \end{align}
The deformation obtained for the dust matter content is quite similar
to the one presented for the scalar field. The five free functions of the
radial component of the triad $\xi_i(\erad)$, with $i=0,1,2,3$, as well
as $\xi_4=\xi_4(\erad,\Phi)$,
encode possible inverse-triad corrections. The formalism also admits
two free functions $\overline{\mathcal{V}}_1(\erad,\Phi)$ and $\overline{\mathcal{V}}_2(\erad,\Phi)$ that depend on the dust field and on the radial component of the triad. The former one
modifies the kinetic energy of the dust field, whereas the latter can be interpreted as a potential
term. In fact, in the particular case this potential acquired a constant value, it would reproduce the cosmological constant.
The classical form of the Hamiltonian (with a vanishing cosmological constant)
is obtained for the case $\xi_i=1$, with $i=0,1,2,3,4$, and $\overline{\mathcal{V}}_1=\overline{\mathcal{V}}_2=0$.

Note that the algebra obtained for this modified Hamiltonian and the one
derived for the scalar-field case are equal and it is given by relations (\ref{eq:classdd}, \ref{eq:poldh}, and \ref{eq:polhh}). Furthermore,
the modified Hamiltonian constraint \eqref{eq:polhamdust} can
also be separated into geometric and matter parts, {\small $\widetilde{\ham}=\widetilde{\ham}_g^{(m)}+\widetilde{\ham}_d$}, with {\small$\widetilde{\ham}_g^{(m)}$} being equal to the modified geometric Hamiltonian \eqref{modhamg} obtained for
the scalar field model. The dust contribution is given by,
  \begin{align}
      \widetilde{\ham}_d &:= \eta\Bigg[\xi_1\sqrt{\xi_0\bar{P}_\Phi^2+(\ephi)^2\mathcal{\overline{V}}_1(\erad,\Phi)}\sqrt{\xi_4+\frac{\erad(\Phi')^2}{(\ephi)^2}}\,+\sqrt{\erad}\ephi\overline{\mathcal{V}}_2(\erad,\Phi)\Bigg]\:.
  \end{align}
 Finally, a brief comment about solutions not contained in the previous derivation, which are also quite similar to the ones presented
  for the scalar-field model. More specifically, we find the following four different families of singular solutions:
 \begin{changemargin}{0.5cm}{0cm}
    \begin{enumerate}[label=(\roman*)]
  \item    $\eta_1=0$, $\eta_3=0$ and $f,g$ do not depend on $P_\Phi$.    \item   $\eta_1=0$, $\eta_4=0$ and $f,g$ do not depend on $\krad$.
    \item   $\eta_1=0$ and $f,g$ do not depend on $\krad$ nor $P_\Phi$.    \item   $\eta_1=0$, $\eta_3=0$ and $\eta_4=0$.
\end{enumerate}
\end{changemargin}

Each of these conditions provides a consistent modification of the classical Hamiltonian constraint when replaced in expression \eqref{genericdusthamiltonian}.
 Once again, the last option corresponds to the homogeneous limit.

\section{Conclusions}\label{sec:conclusions}

In this paper, we have considered possible inverse-triad and holonomy modifications
for spherically symmetric models in vacuum as well as coupled to simple matter models.
For such a purpose, the generic forms \eqref{eq:polhamg}, \eqref{modifiedhamiltonianphi} and  \eqref{genericdusthamiltonian} of the modified Hamiltonian constraint
have been proposed, for vacuum, scalar-field and dust model, respectively.
In these expressions we have left complete freedom
for the dependence on the triad and curvature components;
whereas the radial derivatives of the variables have been
left as in the classical Hamiltonian.
In particular,
we have allowed for non-minimal couplings between
geometric and matter degrees of freedom. Even if the classical model is
minimally coupled to matter, it might happen that, as we approach the quantum
regime, non-minimal couplings are developed. In addition, possible polymerization
of matter fields are also considered in the mentioned modified Hamiltonians.

The modification of the diffeomorphism constraint has only been considered
for the vacuum model. Firstly, the full theory of loop quantum gravity
does not regularize this constraint and, therefore, such corrections are not expected.
Secondly, in App. \ref{app:moddiff} we consider a quite generic deformation
of the diffeomorphism constraint for the vacuum model and show that, under the
requirement of forming a closed algebra, the only allowed modification
is a global multiplicative function, in agreement with results presented in \cite{nodiff}. This points out that the deformation of the diffeomorphism constraint is trivial and
does not provide additional freedom for further corrections
in the Hamiltonian.

The requirement that these modified constraints form a first-class algebra turns out to be quite restrictive. For each model (vacuum, scalar field and dust), the final expression for the modified Hamiltonian (\eqref{eq:polhamgrav}, \eqref{eq:polhamscalargauge} and \eqref{eq:polhamdust},
respectively) has been obtained.
 These modified Hamiltonians are the main result of our study and represent a family of consistent deformations of spherically symmetric general relativity.

Note that, apart from the (trivial) global multiplicative function, none of these Hamiltonians
contain free functions of the angular component of the triad $\ephi$,
which rules out the possibility of including inverse-triad corrections
associated to this component. In addition, only the vacuum
Hamiltonian \eqref{eq:polhamgrav} presents free functions ($h_1$ and $h_2$) of the angular
curvature component $\kang$. Both are related through \eqref{eq:anomalg}
and can be interpreted as holonomy corrections. On the other hand, no
holonomy modifications involving the radial component of the curvature
$\krad$ are allowed, not even in vacuum. Furthermore, the dependence of
the modified Hamiltonian on the momenta of the matter sector
is also bound to its classical form. Hence, no
polymerization of the matter variables is allowed under the present assumptions.

In our analysis we have shown that the
absence of corrections involving $E^\varphi$ and $K_r$ can be understood from the
requirement of mathematical consistency of the effective theory.
Nonetheless, inverse-triad corrections for $E^\varphi$ and holonomy modifications including $K_r$ are not usually considered in the literature with some justification on physical grounds. First, in spherical symmetry, the area of the underlying elementary plaquettes naturally depends on $E^r$ but not on $E^\varphi$ and, thus, inverse-triad corrections are expected to depend only on the radial component of the triad. Second,
for the non-compact radial direction, one would need to consider non-local
holonomy corrections for $K_r$. Although a proposal in this direction (based on a truncated expansion on derivatives of $K_r$) can be found in reference \cite{derivativeexpansion}, the implementation of such non-local modifications remains as an open question.
  
Contrary to the vacuum case, no holonomy corrections are allowed when matter fields are included. This is a direct consequence of the presence of the $\phi'$ and $\Phi'$ terms in the anomaly equations \eqref{eq:anomalhh} and \eqref{eq:anomalhhdust}, respectively. Whereas in the vacuum case the two remaining free functions ($h_1$
and $h_2$) are required to fulfill a unique relation, \eqref{eq:anomalg}, in each matter model an additional restriction arises, \eqref{eq:anodiff2} and \eqref{g0tilde}. This further condition completely fixes the dependence of the
Hamiltonian on the angular curvature component $\kang$ to its classical
form and leaves no room for holonomy corrections.

In addition, we have also found some \emph{singular} solutions (in the sense that they are
not contained in the general solutions leading to the forms \eqref{eq:polhamgrav} \eqref{eq:polhamscalargauge} and \eqref{eq:polhamdust} of the Hamiltonian)
to the anomaly equations that provide an anomaly-free algebra.
These solutions provide
a consistent deformation of the classical Hamiltonian assuming the vanishing of one (or several) correction functions. In these particular scenarios, different radial derivatives of the configuration variables are removed and a higher degree of freedom is acquired for the remaining terms. In particular, holonomy corrections are allowed in each case (see, for instance, the example \eqref{hamiltonianwithholonomies} for a scalar-field matter content). We conclude that the presence of radial derivatives of a given variable
in the Hamiltonian is the main restriction regarding the polymerization of its canonical counterpart.

Finally, let us comment that
the main limitation of our analysis is that no free dependence on
variables with radial derivatives has been allowed in the generic form of the modified
Hamiltonians. Nonetheless, our results point out quite generically to
the impossibility of including holonomy corrections in the presence of
matter fields making use of the real-valued Ashtekar-Barbero variables.
As an alternative, in the last few years, some proposals have
been made
by considering self-dual variables \cite{Achour:2015zmk,selfdualpert,selfdual,selfdual2}.

\begin{acknowledgments}

  AAB acknowledges financial
support from the FPI fellowship PRE2018-086516 of the Spanish Ministry
of Science, Innovation and Universities. This work is funded
by Project \mbox{FIS2017-85076-P} (MINECO/AEI/FEDER, UE)
and Basque Government Grant No.~IT956-16.

\end{acknowledgments}

\appendix

\section{Modified diffeomorphism constraint in vacuum}\label{app:moddiff}

  In this appendix, we will consider a deformed form for the diffeomorphism
constraint in vacuum and demand that it closes algebra along with the deformed
Hamiltonian constraint \eqref{eq:polhamg}. The result will be that the only allowed correction for the diffeomorphism constraint is a global function, which generalizes the previous result presented in \cite{nodiff}.

We will allow for a multiplicative correction function on each term of the diffeomorphism
constraint or, equivalently, a global function $\Omega$ and a correction function $\omega$
on one of the terms. In this way, we start our analysis from the general
form,
  \begin{equation}
      \widetilde{\diff}_g =\, \Omega(\erad,\ephi,\krad,\kang)\,\Big[(\kang)'\ephi-\omega(\erad,\ephi,\krad,\kang)\,\krad(\erad)'\Big]\:.
  \end{equation}
  with both corrections depending on all the four variables.
  
  This diffeomorphism constraint along with the modified Hamiltonian \eqref{eq:polhamg} 
  produces five different anomalous terms:
\begin{align}
\big\lbrace\widetilde{D}_g[\shift_1],\widetilde{D}_g[\shift_2]\big\rbrace&\rightarrow \mathcal{A}_\diff^{DD},\\
\big\lbrace\widetilde{D}_g[\shift],\widetilde{H}_g[\lapse]\big\rbrace&\rightarrow \mathcal{A}_\diff^{DH0},\, \mathcal{A}_\diff^{DH1},\, \mathcal{A}_\diff^{DH2},\\
\big\lbrace\widetilde{H}_g[\lapse_1],\widetilde{H}_g[\lapse_2]\big\rbrace&\rightarrow\mathcal{A}_\diff^{HH}.
\end{align}  
  The anomalous terms $\mathcal{A}_\diff^{DH0}$, $\mathcal{A}_\diff^{DH1}$, and $\mathcal{A}_\diff^{DH2}$ from the second bracket are
  multiplied by $\shift,(\shift)'$ and $(\shift)''$, respectively, and thus they must vanish independently. From now on, we will exclude irrelevant global factors.
  
  Let us first focus on $\mathcal{A}_\diff^{HH}$:
    \begin{align}  
      \mathcal{A}_\diff^{HH} &:=      
      {(\erad)'}\Bigg[\frac{\ephi}{2\erad}\left(\eta_1-\eta_3+2\erad\frac{\partial\eta_1}{\partial\erad}\right)\frac{\partial f_g}{\partial\krad}+\eta_1\left(\frac{\partial f_g}{\partial\kang}-\omega\krad\frac{\partial^2 f_g}{\partial\kang\partial\krad}-\ephi\frac{\partial^2 f_g}{\partial\erad\partial\krad}\right)\!\!\Bigg]\nonumber\\
    &+\ephi(\ephi)'\Bigg[\eta_1\frac{\partial^2 f_g}{\partial\ephi\partial\krad}-\frac{\partial\eta_1}{\partial\ephi}\frac{\partial f_g}{\partial\krad}\Bigg]-\eta_1\ephi(\krad)'\frac{\partial^2 f_g}{\partial\krad^2}\;.     
  \end{align}
   We see that it is very similar to the one found in Sec.~\ref{sec:polgrav}.
   As in that case, we obtain that the form
  \begin{align}\label{eq:fg}
      f_g = \eta_1(\erad,\ephi)\Big[f_0(\erad,\ephi,\kang)+\krad f_1(\erad,\kang)\Big],
  \end{align}
  cancels the coefficients of $(\krad)'$ and $(\ephi)'$. In addition, the vanishing of the anomaly demands:
    \begin{align}\label{eq:conddiff1}
      0&=\,\left(1-\frac{\eta_3}{\eta_1}\right)\frac{f_1\ephi}{2\erad}+\frac{\partial f_0}{\partial\kang}+(1-\omega)\krad\frac{\partial f_1}{\partial\kang} -\ephi\frac{\partial f_1}{\partial\erad},
  \end{align}
  which will be further analyzed below.
  On the other hand, for the anomaly $\mathcal{A}_\diff^{DD}$ we find
  \begin{equation}
      \mathcal{A}_\diff^{DD} \propto \mathcal{A}_\diff^{DH2} -\ephi\frac{\partial\omega}{\partial\ephi}\:,
  \end{equation}
  which means that $\omega$ cannot depend on $\ephi$ as $\mathcal{A}_\diff^{DH2}$ must vanish on its own.
  In fact, the condition $\mathcal{A}_\diff^{DH2}=0$ reads explicitly,
  \begin{equation}
      0=1-\omega-\krad\frac{\partial\omega}{\partial\krad},
  \end{equation}
  which is solved by
  \begin{equation}\label{eq:omega1}
      \omega=1+\frac{\omega_1(\erad,\kang)}{\krad}.
  \end{equation}
  We implement now conditions \eqref{eq:fg} and \eqref{eq:omega1} in the anomaly $\mathcal{A}_\diff^{DH1}$:
  \begin{align}
      \mathcal{A}_\diff^{DH1} &=    
      \frac{\big[(\erad)'\big]^2}{8}\Bigg(\frac{\partial\eta_1}{\partial\ephi}\eta_3-\eta_1\frac{\partial\eta_3}{\partial\ephi}+4\eta_1^2\frac{\erad}{(\ephi)^2}\frac{\partial\omega_1}{\partial\kang}\Bigg)\nonumber\\
    &+\frac{\eta_1^2}{\sqrt{\erad}}\left(f_0-\omega_1 f_1- \ephi\frac{\partial f_0}{\partial\ephi}\right)+\frac{(\ephi)^2}{2}\left(\eta_1\frac{\partial\eta_2}{\partial\ephi}-\frac{\partial\eta_1}{\partial\ephi}\eta_2\right).   
  \end{align}
Without loss of generality, we set $\eta_2=\xi_2(\erad)\eta_1$ as in previous derivations, which annihilates the last term.
Since there is no dependence of the free functions on radial derivatives,
the vanishing of the above anomaly is then equivalent to the following two equations:
  \begin{align}\label{eq:b9}
      0&=\frac{\partial\eta_1}{\partial\ephi}\eta_3-\eta_1\frac{\partial\eta_3}{\partial\ephi}+4\eta_1^2\frac{\erad}{(\ephi)^2}\frac{\partial\omega_1}{\partial K_\varphi},\\
     0&=f_0-\omega_1f_1-\ephi\frac{\partial f_0}{\partial\ephi}.
      \end{align}
In the first equation, the only potential dependence on the angular component of the curvature $\kang$ is contained in
the function $\omega_1$. Therefore it must be at most linear in that variable, $\omega_1:=\xi_d(\erad)+\kang\,\xi_\omega(\erad)$; otherwise
this equation would not be satisfied.
Taking this into account, the respective solutions of the last equations are.:
  \begin{align}
      \eta_3&=\eta_1\left[\frac{\xi_3(\erad)}{\xi_1(\erad)}-4\frac{\erad}{\ephi}\xi_\omega(\erad)\right],\\
 f_0&=\omega_1 f_1 +\ephi f_2(\erad,\kang).
  \end{align}
Plugging these expressions into \eqref{eq:conddiff1} leads to:
  \begin{equation}  
      0=\frac{\partial f_2}{\partial\kang}+\frac{f_1}{2\erad}\left(1-\frac{\xi_3}{\xi_1}+6\frac{\erad}{\ephi}\xi_\omega\right)-\frac{\partial f_1}{\partial\erad}.        
  \end{equation}
Note that the only dependence of this equation on $\ephi$ is explicit, as none of the free functions depends on it.
This means that either $f_1=0$ or $\xi_\omega=0$ should be obeyed. The first option is not of interest since it would further
imply that $f_2$ is independent of $\kang$ and one would then obtain that the function $f_g$ (and thus the modified Hamiltonian) would be completely
independent of the curvature components $\krad$ and $\kang$. As we are not considering singular solutions, we impose $\xi_\omega=0$.
Finally, defining {\small$h_1:=-\xi_1f_1/(2\sqrt{\erad})$} and {\small$h_2:=-2\sqrt{\erad}\xi_1f_2$}, with $\xi_1=\xi_1(\erad)$, the last
relation takes the same form as the anomaly equation \eqref{eq:anomalg}.

After imposing all these conditions, the last anomaly $\mathcal{A}_\diff^{DH0}$ also vanishes
and the modified diffeomorphism constraint reads,
  \begin{equation}
      \widetilde{\diff}_g = \Omega(\erad,\ephi,\krad,\kang)\,\Big[(\kang)'\ephi-\Big(\krad+\xi_d(\erad)\Big)(\erad)'\Big]\:.
  \end{equation}
The function $\xi_d(\erad)$, also present in the Hamiltonian constraint, can be absorbed performing the canonical transformation
$\bar{K}_r=K_r+\xi_d$. In this way, the only modification to the diffeomorphism
constraint is a global multiplicative function $\Omega$. As commented in
the main body of the article, this kind of global multiplicative functions are trivially allowed in the constraints by the requirement of anomaly freedom.
 
Finally, the modified Hamiltonian takes the same form \eqref{eq:polhamgrav} as above (with the variable $\bar{K}_r$ instead of the initial $\krad$) and, therefore,
we conclude that allowing for modifications to the vacuum diffeomorphism constraint
does not involve additional freedom to include more corrections in the
Hamiltonian constraint.

\section{Leading-order holonomic corrections in the anomaly-free semiclassical sector}\label{app:leading}

    In this appendix, an approximate solution to the equation \eqref{eq:anomalg} will be found, which
  will provide the leading-order terms of the holonomy corrections in the pure vacuum model.
  Following the notation of the improved-dynamics scheme of loop quantum cosmology,
  we include a parameter $\bar{\mu}=\bar{\mu}(\erad)$ that accounts for the discreteness scale.
  Note that if the holonomy corrections $h_1$ and $h_2$ are scale-independent,
  both functions are allowed to be periodic in $\kang$, as the last term in \eqref{eq:anomalg} drops out.
  In contrast, when one considers a length-dependent holonomy
  $h_1$, the additional requisite that $\partial h_1/\partial\erad$ is periodic in $\kang$ must also be satisfied.  For example, a sinusoidal function as $\holbar=\sin{(\bar{\mu}\kang)}/\bar{\mu}$, with $\bar{\mu}={\bar\mu}(\erad)$,
  produces a term proportional to $\kang$ on the right-hand side of \eqref{eq:anomalg}, and thus $h_2$ can not be periodic in $\kang$.
  However, as we are working in a semiclassical approach, we can focus on leading-order holonomy corrections and
  try a power-series solution for $h_1$ and $h_2$. Let us assume that both holonomy corrections can be written as follows:
  \begin{align}
  h_{1}&= \sum_{n=0}^\infty\, a_n\, \bar{\mu}^n\,{\kang}^{n+1}\:,\\
  h_{2}&= \sum_{n=0}^\infty\, b_n\, \bar{\mu}^n\,{\kang}^{n+2}\:,
  \end{align}
  where the coefficients $a_n$ and $b_n$ are allowed to depend on $\erad$.
  Assuming now the form $\bar{\mu}(\erad)=(\Lambda^2/\erad)^{N/2}$, with a constant $\Lambda$, the anomaly \eqref{eq:anomalg} provides the relation:
  \begin{align}\label{eq:relation}
    \frac{n+2}{2} b_n = \left(\frac{\xi_3}{\xi_1}-2\erad\frac{\partial\log{\xi_1}}{\partial\erad}-nN\right)a_n +2\erad\frac{\partial a_n}{\partial\erad}.
  \end{align}
  If we demand now constant values for the coefficients $a_n$ and $b_n$ (as one would expect if $\bar{\mu}$ retained all the scale dependence), and no inverse-triad corrections ($\xi_1=\xi_3=1$),
  the last term in \eqref{eq:relation} drops out. In a sinusoidal approximations for $\holbar$, only even coefficients would survive and therefore,
  \begin{equation}
    b_{2n} = -\frac{2n-1}{n+1}a_{2n}\:,
  \end{equation}
  where we have chosen $N=1$. For instance,
  if the usual ansatz $\holbar=\sin{(\bar{\mu}\kang)}/\bar{\mu}$ is chosen, the leading terms
  for the holonomy correction functions are explicitly given by,
  \begin{align}
      \holbar &\approx \kang -\frac{1}{6}\frac{\Lambda^2}{\erad}\kang^3\:,\\
      \hol &\approx \kang^2 +\frac{1}{12}\frac{\Lambda^2}{\erad}\kang^4\:.
  \end{align}

\end{document}